\documentclass[useAMS,usenatbib,usegraphicx]{mn2e}

\usepackage{amsmath}
\usepackage{amssymb}

\newcommand{\aap}{A\&A}
\newcommand{\aaps}{A\&AS}
\newcommand{\actaa}{AcA}
\newcommand{\aj}{AJ}
\newcommand{\araa}{ARA\&A}
\newcommand{\apj}{ApJ}
\newcommand{\apjl}{ApJ}
\newcommand{\apjs}{ApJS}
\newcommand{\apss}{Ap\&SS}
\newcommand{\baas}{BAAS}
\newcommand{\mnras}{MNRAS}
\newcommand{\nat}{Nature}
\newcommand{\pasp}{PASP}

\title[Discovery of $\sim$ 9,000 new RR Lyrae in the South] {Discovery
  of $\sim$ 9,000 new RR Lyrae in the Southern Catalina Surveys}

\author[G. Torrealba et al.]
{G.~Torrealba$^{1,2}$,
M.~Catelan$^{1,3}$, 
A.J.~Drake$^4$, 
S.G.~Djorgovski$^4$,
R.H. McNaught$^5$,
\newauthor
V.~Belokurov$^2$,
S.~Koposov$^2$, 
M.J.~Graham$^4$,
A.~Mahabal$^4$,
S.~Larson$^6$,
\newauthor
and E.~Christensen$^5$ \\
$^1$Instituto de Astrof\'{i}sica, Pontificia Universidad Cat\'olica de Chile, Av. Vicu\~na Mackena 4860, 782-0436 Macul, Santiago, Chile\\
$^2$Institute of Astronomy, Madingley Rd, Cambridge, CB3 0HA\\
$^3$Millennium Institute of Astrophysics, Santiago, Chile \\
$^4$California Institute of Technology, 1200 E. California Blvd, CA 91225, USA\\
$^5$The Australian National University, Siding Spring Observatory, Coonabarabran, NSW, Australia\\
$^6$The University of Arizona, Department of Planetary Sciences, Lunar and Planetary Laboratory, 1629 E. Tucson AZ 85721, USA
}

\begin{document}

\date{Accepted for publication in MNRAS 2014 October 28. Received 2014 October 8; in original form 2014 March 27}

\pagerange{\pageref{firstpage}--\pageref{lastpage}} \pubyear{2014}

\maketitle

\label{firstpage}

\begin{abstract}
We present the results of a deep, wide-area variability survey in the
Southern hemisphere, the first of its kind. As part of the Catalina
Sky Surveys, the Siding Spring Survey (SSS) has covered $14,\!800$
square degrees in the declination range of
$-75^{\circ}\leq\delta\leq-15^{\circ}$. To mine the enormous SSS
dataset efficiently we have developed two algorithms: Automatic Period
Selection (APS) and Automatic Fourier Decomposition (AFD), which aim
to sharpen the period estimation and produce robust lightcurve
models. Armed with the APS and AFD outputs we classify $10,\!540$
ab-type RR~Lyrae (RRab) stars ($\sim$90\% of which are new) across the
Southern sky. As well as the positional information we supply
photometric metallicities, and unreddened distances.

For the RRab stars in the halo, a study of the photometric metallicity
distribution reveals a nearly Gaussian shape with a mean metallicity of ${\rm
[Fe/H]}=-1.4$~dex and a dispersion of $0.3$~dex. A spatial study of the RRab
metallicities shows no significant radial gradient in the first $\sim7$~kpc
from the Galaxy center. However, further out, a small negative gradient is
clearly present. This is complemented by a very obvious correlation of the
mean RR Lyrae metallicity with distance above the Galactic plane, $z$. We have
also carried out an initial substructure search using the discovered RRab, and
present the properties of the candidates with significance greater than
$2~\sigma$. Most prominent among these is a southern extension of the Sagittarius 
dwarf galaxy's stream system, reaching down to declinations $\sim -40\deg$.

\end{abstract}


\section{Introduction}\label{sec:INTRO}

The search is on to identify the fragments of those dwarf satellites
that fell into the Milky Way many Gyrs ago. Even the most massive of
these unfortunate galaxies must have had their star-formation
throttled rather rapidly as the halo appears to be dominated by the
old and metal-poor population \citep[see
  e.g.][]{Tolstoy2009}. Scattered over hundreds of kiloparsecs, these
low surface brightness tidal debris are outnumbered by the much closer
disc and bulge stars with a ratio of at least 100:1. Luckily, the halo
archetypical constituent, the RR Lyrae are easily recognized even at
faint magnitudes provided that their lightcurves are measured with
sufficient precision. Indeed, these old pulsating horizontal-branch
stars have long been used as fundamental distance probes, shedding
flickering light on the history of the Galaxy formation
\citep[e.g.,][and references therein]{mc09}.

Within the generally accepted $\Lambda$CDM paradigm, the formation of
the stellar halo reflects the evolution of the shadowy one, made of
dark matter. Both are assembled through the action of gravity mostly,
but the links between the two are complex and their subtleties are
only starting to be explored \citep[e.g.][]{Deason2013}. Even if the
``uniqueness'' of the Milky Way is often viewed as both virtue and
vice, its place as a laboratory to study the structure formation in
the low-mass regime is not contested. Only here, at home, the fine
details of the hierarchical processes driven by the gravitational
forces of the large-scale distribution of cold dark matter can be
scrutinized with ample resolution and rigor \citep[see
  e.g.][]{kfjb02}. As shown by both data and simulations
\citep[e.g.][]{Weisz2014, Brooks2014}, the lives of the accreted
dwarfs are always severely affected. Even if some of them survive,
their gas is stripped and as a result their star-formation hindered or
altogether put out. Recently, such cinders of ancient star-formation
have been uncovered within the Galactic halo in large numbers
\citep{Willman2010, BelokurovReview2013}. While it is still not
possible to compare directly the properties of the observed and the
simulated dwarf satellites, it appears that the number of puzzling
discrepancies is slowly growing \citep[see
  e.g.][]{Klypin1999,mbmr11,mbea11}.

It is predicted that if the Galactic tides are strong enough, the
infalling dwarf satellites are pulled apart leaving trails of stars
that can be recognized as coherent overdensities in the halo
\citep[e.g.,][]{kjea96,phea01}. With the advent of deep all-sky
surveys, the examples of prehistoric and ongoing disruption events
have been seen all around the Galaxy
\citep[e.g.,][]{kvea01,Newberg2002,smea03,fos2006,cg06,hac2007,orphan2007,adea13a}. Despite
the abundance of the Galactic tidal debris detections, a detailed
quantitative comparison with state-of-the-art numerical simulations
has not yet been carried out. Curiously, the few studies that do
attempt to gauge the lumpiness of the Milky Way stellar halo in both
real and mock datasets, tend to conclude that the Galaxy is somewhat
smoother than predicted \citep[see e.g.][]{Helmi2011,
  Deason2011}. This might simply be because the total stellar mass in
the myriad of the recently discovered sub-structures is relatively
small. Such estimates of the halo lumpiness rely of course on the
knowledge of the global properties of the stellar halo, including its
integrated luminosity and overall extent. These, however, are poorly
determined, as large portions of the Galaxy remain unexplored: the
most obvious lacunae being the low-latitude inner portions of the halo
as well as large swathes of the Southern sky.

The Catalina Sky Survey has collected multi-epoch imaging data from three
different observing sites, covering most of the accessible sky, namely
the area from $\delta=-75^{\circ}$ to $\delta=+65^{\circ}$. The
Survey's results in the Northern hemisphere have been presented
recently \citet{adea13a,adea13b}. Here, we analyse data from the
Southern hemisphere segment of Catalina, the Siding Spring Survey
(SSS). The aim of our analysis is to fill in the obvious gap in the
South, and discover and characterize ab- type RR Lyrae stars with
$-75^{\circ}<\delta<-15^{\circ}$. RR Lyrae stars, and in particular those of
the ab-type (RRab), have long been known as a trusted tracer of the
Galactic halo. The RRab's are relatively easy to differentiate from
other types of variable stars due to the characteristic asymmetric
shape of the lightcurve. We, therefore, concentrate solely on RRab's
which allows us to assemble a comparatively clean sample of stars over
$\sim 14,\!800$ square degrees. As a result, when combined with the
previously published Catalina Sky Survey catalogues, our total RR
Lyrae sample covers more than $33,\!000$ square degrees and reaches
distances in excess of 50 kpc.

Thanks to the immense, un-interrupted spatial extent of our survey, we
can, for the first time, build a global inventory of the halo
sub-structures. Most importantly, it is now possible to quantify
large-scale asymmetries \citep[e.g.][]{vpos} if any are indeed
present. Recently, several bloated halo structures covering hundreds
of square degrees have been identified in the Northern sky, including
the Hercules-Aquila Cloud \citep[HAC;][]{hac2007} and Virgo
Overdensity \citep[VOD;][]{Vivas2008,mjea08}. In their morphology, these stellar
``clouds'' are distinct from narrow tidal tails or ``streams'', and
were likely produced as a result of nearly radial in-fall of
relatively massive galaxies, but their exact origin is still
unknown. With the data from one half of the sky missing, it has been
impossible so far to establish whether several large inner halo
structures are connected in any way. Given that both Virgo \citep[see
  e.g.][]{Duffau2014} and HAC \citep{Simion2014} contain copious
amounts of RR Lyrae, we hope that our new sample will help to complete
the puzzle.

Finally, we also hope to leverage thousands of precise distances
across the Southern sky to understand better the most prominent
un-relaxed structure in the outer Galaxy, the Sagittarius stream. The
Sagittarius dwarf spheroidal galaxy \citep{riea94} and its associated
tidal tails \citep[see e.g.][]{smea03} are probably the best-studied
example of the Galactic halo sub-structure. However, after two decades
and dozens of published papers \citep[see
  e.g.][]{Jiang2000,madf2006,dlsm10,Yanny2009,Koposov2012,sgrprec}
there remain many un-answered questions about the Sgr accretion
event. The Sgr stream is so long that its wraps overlap in many
locations on the sky thus confusing the interpretation. Our data
possesses the necessary accuracy to resolve the sub-structure along
the line of sight, therefore helping to untangle Sgr's intertwined
tidal tails.

The main objective of this paper is to automate the RR Lyrae catalog
production. Therefore, we start by introducing the survey data in
Section~\ref{sec:ODC} and then go on to describe the algorithms
designed to find and classify the RRab in
Section~\ref{sec:SELECTION}. In Section~\ref{ssesc:CLASS} we give the
details of the calculation of the RRab fundamental
parameters. Finally, the properties of the discovered RRab stars are
presented in Section~\ref{ssec:RES}.

\section{Data and Calibration}\label{sec:ODC}

In this paper we analyze 18,288 periodic variable candidates found in
the SSS, the Southern part of the Catalina Sky Survey. The Catalina
Sky Survey began in 2004 and utilized three telescopes at three
different observing sites to cover the sky between declination
$-75^{\circ}<\delta<+65^{\circ}$. Acquiring images of the sky across
more than 33,000 square degrees, the principal aim of the Catalina Sky
Survey was to discover Near-Earth Objects (NEOs) and Potential
Hazardous Asteroids \citep[PHAs;][]{slea03}. Its counterpart, the
Catalina Real- time Transient Survey (CRTS), involves the analysis of
the same data in order to detect and classify stationary optical
transients \citep{adea09}. Both surveys work collaboratively to
extract the maximum scientific return from the data of the three
telescopes operated by the Catalina Sky Surveys. These consist of the
Catalina Schmidt Survey (CSS) telescope, the Mount Lemmon Survey (MLS)
telescope, and the SSS telescope in Siding Spring, Australia.  Each
telescope is equipped with a 4k$\times$4k CCD camera, with the field
of view of 8.2, 1.1 and 4 square degrees for the CSS, MLS and SSS,
respectively.  The focus of this work, the SSS uses the Uppsala 0.5-m
Schmidt telescope equipped with a CCD camera nearly identical to the
one used by its northern counterpart, the CSS, which is an obvious
advantage in terms of homogeneity of the resulting photometry. The
pixel size is 2 arc minutes giving a field of view of 4 square
degrees. In the SSS observations, the Galactic plane region
($|b|<15^{\circ}$) is avoided due to crowding. Images are unfiltered
to maximize throughput, and taken in sequences of four observations
separated by ten minutes with exposures that last approximately 30
seconds. SExtractor \citep{ebsa96} is employed to obtain photometry of
individual objects in an automated fashion.

\begin{figure*}
    \includegraphics[width=168mm]{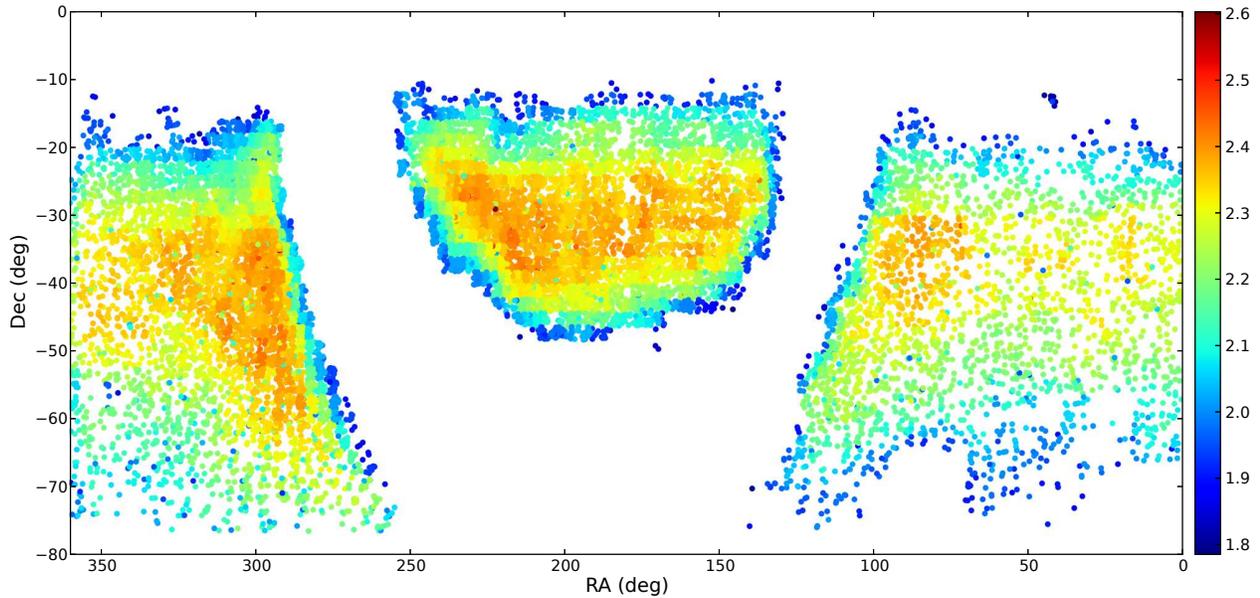}
    \caption{Spatial extent of the SSS sample in equatorial
      coordinates. Each filled circle is a candidate periodic variable
      as selected by the Catalina pipeline. The color represents the
      logarithm of the number of observations, as indicated by the
      scale on the right. Note the drop in frequency of visits close
      to the edges of the survey.}
    \label{fig:fig01}
\end{figure*}

The initial selection of variable objects and the subsequent
photometric transformations are all part of the general Catalina
pipeline and are identical to those described in \citet{adea13a}, and
therefore are only briefly summarized here. First, variable source
candidates were identified from the SSS data using the Welch-Stetson
variability index $I_{ws} > 0.6$ \citep{WelchStetson1993}. Then, for
all objects in the list of potential variables, the Lomb-Scargle
period finding method \citep{nl76,js82} was applied and periodic
variable candidates with periods between 0.1 and 4 days were
selected. These boundaries are chosen to be broader than the expected
range of periods for RRab stars ($\sim 0.4 - 1$~days) to make sure
that we also recover RRab stars affected by aliasing in the
Lomb-Scargle analysis. The sample of periodic variables consists of
$\sim20,\!000$ objects with unfiltered magnitudes between 10.5 and
19.2 and $\sim100$ epochs each. Figure \ref{fig:fig01} shows locations
of all candidate periodic objects in equatorial coordinates, to give
an idea of the spatial extent of the sample.


Since the CCD used by the SSS matches closely the one used by the CSS
(implying, of course, that the responses are nearly identical), we can
apply the calibration equations implemented by the CSS to transform
between the unfiltered Catalina SSS data and the standard Johnson $V$
filter. This is given in equation 1 of \citet{adea13a}:

\begin{equation}
V=V_{SSS}+0.31(B-V)^{2}+0.04.
\label{eq:INTRO1}
\end{equation}

\noindent This relation has a dispersion of $\sigma = 0.059$
mag. Since in this work we focus on RRab stars, we investigate the
impact of color variations on the recovered properties of these
variables. If we consider an average color for RRab stars of $(B-V)
\sim 0.3$ mag and the possible color range of $0.1\,\rm{mag} \la (B-V) \la 0.5\,\rm{mag}$ 
\citep{gkea05}, the maximum difference in color is $\sim 0.2$ mag in
$(B-V)$. This adds an extra uncertainty of $\sim 0.05$ mag to
eq. \ref{eq:INTRO1}, which combined in quadrature with the dispersion
of the relation gives a total dispersion of 0.077 mag in the $V$
magnitude. However, because RRab's change color during their pulsation
cycle, taking the average color might cause measurable effects to the
lightcurve shape, which in turn could lead to a bias in the estimation
of the metallicity and the average magnitude of the star. We tested
these effects by making a simple model of the RRab color change as a
function of the lightcurve phase, following the typical color
variations of the RRab during their pulsation cycles as seen in Figure
2 of \citet{gkea05}. We used this model to transform the $V_{SSS}$ mag
to $V$ mag as a function of phase and estimate the metallicity and the
average magnitude. We then compared these quantities with the one
obtained by taking a constant value of $(B-V) = 0.3$ mag, and found that
the differences have a dispersion of $0.05$ dex for the metallicity
and $0.004$ mag for the average magnitude. These values are an order
of magnitude smaller than the typical uncertainty induced by
photometric error. Since we lack the color information in the SSS
data, and given that the effects of the color variation over the
pulsation cycle are small, we believe that we are justified in using
an average color of $(B-V) = 0.3$ mag to transform between $V_{SSS}$ and
Johnson $V$.

\section{Period Determination}\label{sec:SELECTION}

Efficient detection of RRab in a dataset like SSS is possible solely
thanks to their characteristic lightcurve shapes. However, the
asymmetry in the RRab pulses only reveals itself if the lightcurve is
phased (folded) with the correct period. Moreover, the metallicity and
the intrinsic luminosity of such a variable star both require accurate
lightcurve morphology measurements. Getting the period right is
therefore absolutely crucial for both the identification of RR Lyrae
as well as the subsequent inference of their properties. We tackle the
problem of automating the RR Lyrae discovery and characterization by
developing two routines, the Automatic Fourier Decomposition (AFD),
which builds robust lightcurve models, and the Automatic Period
Selection (APS), which makes use of different periodograms to home in
on the correct period. With these tools in hand, the process of the
classifying variables based on the shapes of their lightcurves is
fast, robust and reproducible. In sections \ref{ssec:AFD} and
\ref{ssec:APS} the two routines are introduced and discussed in
detail.


\subsection{Automatic Fourier Decomposition}\label{ssec:AFD}

The essence of the AFD method is that the complexity of the lightcurve
model is dictated by the data's signal-to-noise ratio as well as the
details of the lightcurve morphology. In other words, the number of
harmonics in the Fourier decomposition of a phased lightcurve is
increased as long as the fit continues to be statistically
significant. To find the parameters of the harmonics in the Fourier
decomposition we carry out a weighted least-squares fitting procedure
where the model lightcurve is described by:

\begin{equation}
f(x)= A_{0}+\sum\limits_{n=1}^{n_{max}} A_{n}\sin{\left( \frac{2n\pi t}{P}+\phi_{n} \right)},
\label{eq:AFD1}
\end{equation}

\noindent where $A_{n}$ are the amplitudes, $\phi_{n}$ the phases, $P$
the period and $n_{max}$ the number of harmonics in the
decomposition. To determine the value of $n_{max}$, we perform several
fits with increasing number of harmonics until it is not statistically
significant to add an extra harmonic. The process starts with just one
harmonic:

\begin{equation}
f(x)= A_{0}+A_{1}\sin{\left( \frac{2\pi t}{P}+\phi_{1} \right)}.
\label{eq:AFD2}
\end{equation}

\begin{figure*}
    \includegraphics[width=184mm]{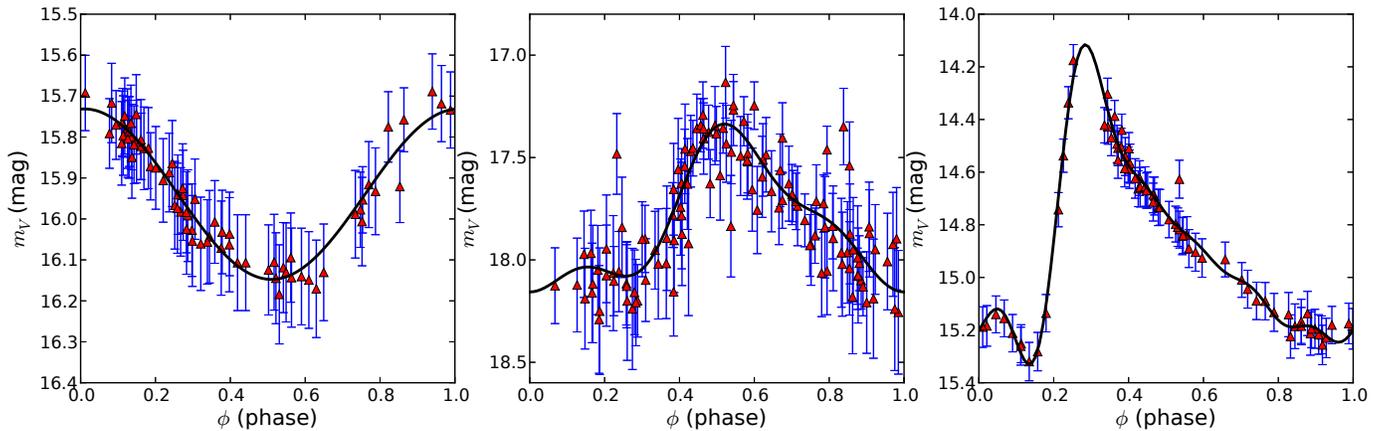}
    \caption{Illustration of the AFD lightcurve fitting (see text for
      more detail) . This shows three examples of phased lightcurves
      (red triangles show the SSS photometric data, while the
      associated error-bars are shown in blue) of three different
      periodic variables as well as their AFD models (black
      curves). From Left to Right: examples of lightcurves that are
      well fit with two (left), four (middle), and six (right) Fourier
      harmonics, respectively.}
    \label{fig:fig1}
\end{figure*}

\noindent We perform a least-squares fit to determine $A_{0}$,
$A_{1}$, and $\phi_{1}$, and quantify the goodness of the fit through
the reduced $\chi^{2}$ statistic, calculated as follows:

\begin{equation}
\chi^{2}= \frac{1}{N-n-1}\sum\limits_{k=1}^{N}\left( \frac{O_{k}-E_{k}}{\sigma_{k}} \right)^{2},
\label{eq:AFD3}
\end{equation}

\noindent where $N$ is the number of observations, $n$ the number of
parameters of the Fourier decomposition, $\sigma$ the error of the
observation, $O$ the observed value, and $E$ the expected value as
given by the Fourier series. We then compare the quality of this fit
to a Fourier decomposition with one extra harmonic and with two extra
harmonics. To determine if the series with extra harmonics gives us a
significant improvement on the quality of the fit, we perform a
statistical F-test to check if the null hypothesis (i.e., the two fits
are statistically equal in terms of their performance) can be
discarded. The $F$ statistic is calculated as follows:

\begin{equation}
F=\left( \frac{\chi^{2}_{1}}{\chi^{2}_{2}}-1\right)\frac{b}{a},
\label{eq:AFD4}
\end{equation}

\noindent where $b=N-p2$ is the number of degrees of freedom of the
second fit, and $a=p2-p1$ is the difference in the number of
parameters $p$ between the second and the first fit (note that $p2>p1$
always). The test statistic is then compared to a critical value,
$\alpha$; if $F$ is less than the critical value, the null hypothesis
cannot be discarded and the two models are statistically equal in
their ability to describe the observed data. To determine $\alpha$, we
require that the rejection probability is $0.99$, that is, the null
hypothesis is discarded when we are 99\% sure that adding a new order
to the series is statistically significant. In addition to this, we
set the maximum number of harmonics to 6, to avoid over-fitting. In
Figure \ref{fig:fig1} we show three lightcurves that were fit by the
AFD with different number of harmonics. It is reassuring to see that
in each case, the algorithm appears to have set the series truncation
according to the complexity of the lightcurve.



\subsection{Automatic Period Selection}\label{ssec:APS}

\begin{figure}
    \includegraphics[width=84mm]{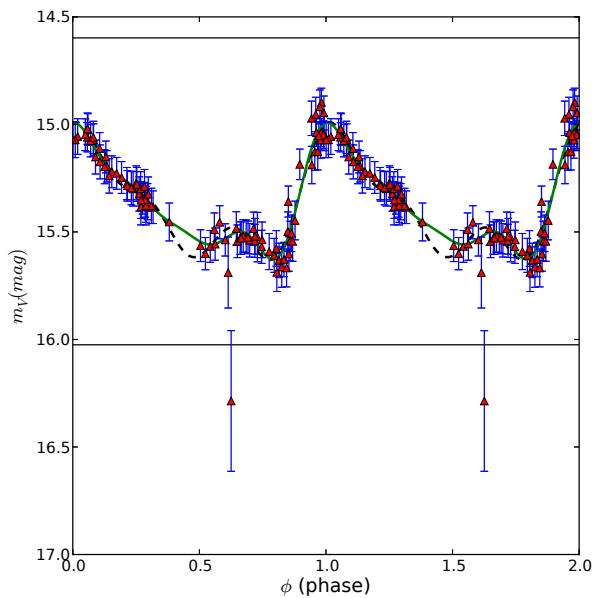}
    \caption{Impact of the first sigma clipping on model
      selection. The dashed black line represents the AFD model
      without sigma clipping, while the green line shows the AFD fit
      after sigma clipping. Horizontal black lines give the threshold,
      $3\sigma$ away from the mean, outside which the data is
      clipped. The impact of the outlier present at $\phi\sim0.1$ is
      clearly visible as the model shown by black dashed line is
      ``pulled'' towards the discrepant datapoint.}
    \label{fig:fig2}
\end{figure}

APS makes period estimation given the available multi-epoch data more
robust. Our approach is to compare periods singled out by different
period finding algorithms and quantify the goodness of fit by applying
the AFD lightcurve model (see Section~\ref{ssec:AFD}). In particular,
we compare the periods proposed by two algorithms, namely Analysis Of
Variance \citep[AOV;][]{as89,as96,as06} and Lomb-Scargle Periodogram
\citep[LS;][]{nl76,js82}. The comparison is done by phasing the
lightcurve with the five best periods proposed by each routine and
calculating the goodness-of-fit parameter to discriminate between the
periods and converge on the best one. Since the SSS photometry
occasionally has some spurious measurements, we apply a simple
cleaning routine by removing the most obvious outliers. In order to do
this, we find the mean magnitude and the overall amplitude of the
lightcurve from an initial AFD and remove all datapoints that lie more
than $3\sigma$ from the mean, where $\sigma$ is the standard deviation
of the mean lightcurve magnitude. With the outliers removed we then perform
a second AFD run. Figure \ref{fig:fig2} displays both the first AFD
model as well as the result of the second AFD run after the outliers
(i.e. datapoints above the sigma-clipping limits shown) have been
removed.


\begin{figure}
    \includegraphics[width=84mm]{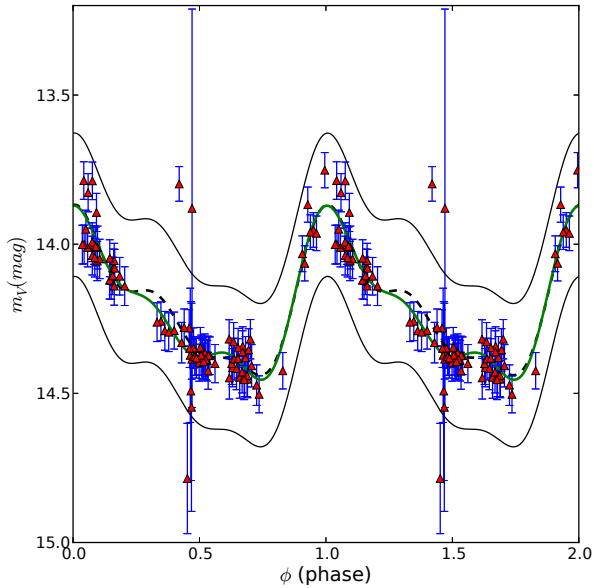}
    \caption{Same as Figure \ref{fig:fig2}, but for a different method
      of clipping. Here, instead of discarding all datapoints that are
      outside a fixed range given by the mean and the standard
      deviation of the lightcurve measurements, a sliding threshold is
      used. At each phase, the clipping threshold is calculated with
      respect to the value of the previous AFD model fit. The effect
      of the datapoints outside the accepted region is less important
      than in the aforementioned plain sigma-clipping process, but
      these can add noticeable wriggles to the AFD fit as illustrated
      by the differences between the black dashed line (before clipping)
      and the green solid line (after clipping).}
    \label{fig:fig3}
\end{figure}

\begin{figure}
    \includegraphics[width=84mm]{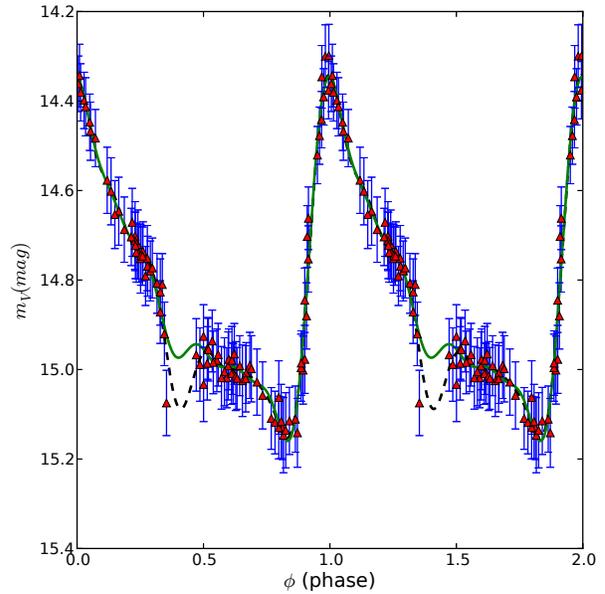}
    \caption{Illustration of the gap filling procedure. Red triangles
      show the SSS photometry at each phase for a chosen period, and
      the associated errors are in blue. Black dashed (green solid)
      line displays the AFD model before (after) the gap filling, As
      with lightcurve clipping, the effect of this procedure is
      secondary in comparison to the raw sigma clipping process, but
      removes considerable wriggles from the AFD.}
    \label{fig:fig4}
\end{figure}

Additionally, there are two extra processes to further clean the
lightcurve if need be: the conventional outlier clipping and the
so-called gap-filling.  Note, however, that the clipping algorithm
does not work well on lightcurves of deep eclipsing binaries, but can,
in principle, be applied to RR Lyrae variables.  The process consists
of clipping all datapoints of the lightcurve that lie more than
$3\sigma$ from the recently computed AFD model at a given phase,
rather than from the overall mean of the lightcurve. In this case,
$\sigma$ is the mean error of the observations. Figure \ref{fig:fig3}
shows the lightcurve clipping boundaries and how these can be used to
excise some of the outliers that have survived the first cut. The
gap-filling process is triggered when there is a gap of more than 0.1
in the lightcurve phase. When this condition is satisfied, a ``data''
point is included in the lightcurve based on the previously calculated
AFD. The synthetic observation is positioned according to a simple
linear interpolation between the phase values at the edges of the
gap. The interpolation points are chosen as the mean of the three
points closest to the edges on each side of the gap (in phase
space). This process relies on the previous AFD run to assign the
appropriate magnitudes around the edge. As seen in Figure
\ref{fig:fig4}, this helps to remove extreme wriggles from the model
lightcurve. The post-processing techniques described above are used to
improve the lightcurve shape estimation when applying the AFD.
Subsequently, we utilize the AFD goodness-of-fit to discriminate
between possible test periods.


%


To select the best period the following rules are applied. First, if
two period candidates have a difference of more than 10\% in reduced
$\chi^{2}$, the best period is selected. If this condition is not met,
but the two first candidates of AOV and LS are similar (i.e. their
difference is lower than $10^{-3}$ days), and match one of the two
best periods ranked by the reduced $\chi^{2}$ test, that period is
selected. If neither of the two criteria is met, but the first two
periods are aliases (either double or half the period), then, since
the AOV periodogram is able to damp the detection sensitivity for
aliased periods \citep{as96}, the one ranked higher by AOV is
chosen. Finally, if neither of these conditions is met, the star is
flagged as having a problematic period. We also flag stars with
reduced $\chi^{2}>3$ and those with periods greater than 4 days, since
given the pre-selection condition mentioned in \S\ref{sec:ODC}, we do
not expect to find stars with periods that long, and they are usually
associated with spurious period candidates. The $\chi^{2}$
distribution of the accepted lightcurves peaks at $\sim 0.3$ rather
than $1$. There are several possibilities to explain the location of
the peak. First, the error of the SSS observations might be
overestimated. Second, we might be over-fitting the lightcurves, and
finally, of course, there could be a combination of the first
two. While overfitting is difficult to avoid in some cases, we point
out that the main point of the AFD is to avoid this and provide better
lightcurve characterizations at the same time. Therefore, we believe
that the peak value of the $\chi^2$ distribution might be due to a
combination of an overestimation of the errors of the observations,
and somewhat over-zealous removal of the outliers which are, of
course, the main causes of high $\chi^2$.

Finally, we test the performance of the APS algorithm on $\sim 1000$
randomly chosen phased lightcurves. Visual inspection of this sub-set
confirms that the method can correctly select the periods when both
AOV and LS agree, when they do not agree, and even when they both fail
with their most significant period candidate. Note that, in order to
avoid contamination, we have made this method very restrictive. In
other words, if the $\chi^2$ test cannot tell the difference between
the two best period candidates, we leave the star out of the
sample. This culls only of order of $6 \%$ of the sample. Combined
with the stars that could not be modeled with the AFD (i.e., reduced
$\chi^{2}>3$), and the stars with periods in excess of 4 days, the
total number of rejected lightcurves is of order of $10 \%$. This is a
reasonably small price to pay, provided we achieve reliable period
estimation for the remaining $\sim 90 \%$ of the stars. We believe,
based on the test described above, that out of these only $\sim 1 \%$
have wrong periods.


\section{Selection of RRab. Distance and metallicity inference}\label{ssesc:CLASS}

\subsection{The M Test}
\begin{figure}
    \includegraphics[width=84mm]{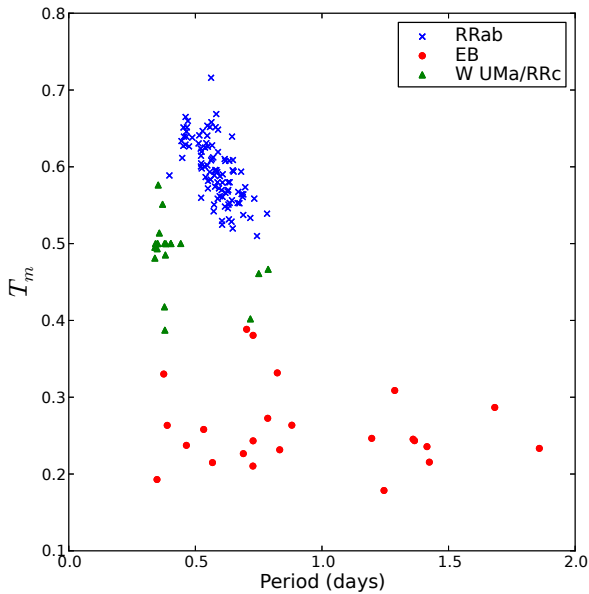}
    \caption{M-statistic as a function of the variable star period for
      a small sub-sample of objects inspected visually. Blue crosses
      are confirmed RRab stars, green triangles are possible RRc
      and/or W~UMa variables, while red filled circles are eclipsing
      binaries (excluding W~UMa systems). This figure clearly
      demonstrates how this simple lightcurve symmetry statistic helps
      to discriminate between the RRab stars and the rest of the
      variables.}
    \label{fig:fig6}
\end{figure}

Armed with the period and the lightcurve shape information, we proceed
to select the most likely RRab stars from the sample of variable
objects. Note that all objects flagged to have unreliable periods (as
described in the previous sections) are removed, which leaves a
total number of $16,\!500$ candidates.  Within the period range occupied by
RRab's, the other common periodic variables are eclipsing binaries. To
differentiate between pulsating and eclipsing stars, we introduce the
M-test, a simple lightcurve shape statistic motivated by
\citet{kkea06}, and already utilized in \citet{adea13a,adea13b}. This
is defined as follows:

\begin{equation}
T_{m}=\frac{M_{max}-M_{mean}}{M_{max}-M_{min}},
\label{eq:CLASS1}
\end{equation}

\noindent where M denotes magnitude. In essence, the M-statistic $T_m$
measures the amount of time the star spends above ($T_{m}>0.5$) or below
($T_{m}<0.5$) a baseline defined by the mean magnitude. To mitigate the
destructive action of outliers, all parameters used to calculate $T_m$ are
extracted from the AFD model. This is the principal difference with the
original method as proposed by \citet{kkea06}, where the actual lightcurve
data is used to calculate the M-statistic. Different types of variables occupy
different positions, in a period vs. $T_m$ diagram. To determine the exact
boundaries of the different classes in this plane, $\sim150$ randomly selected
variables were classified by eye into three groups: RRab's, detached eclipsing
binaries and RRc/W UMa- type eclipsing binaries. Figure \ref{fig:fig6} shows
the locations of these groups in period-$T_m$ space. As is obvious from this
figure, RRab's form a tight and fairly isolated group with high values of the
M-statistic. To further clean our sample, we removed all lightcurves that have
more than one peak in phase space. This is achieved by tagging all the stars
whose amplitude in the higher harmonics of the AFD model is larger than the
amplitude of the first harmonic. With regards to the RR Lyrae of c type, there
possibly exists a region in Figure \ref{fig:fig6}, i.e. at low periods and
intermediate $T_m$, which is predominantly populated by these pulsators.
However, the levels of contamination with W UMa stars are still high.
Accordingly, in order to have the simplest and the cleanest selection
possible, we do not attempt to disentangle RRc's and W UMa's, and focus on
RRab stars only. Table \ref{tab:RRabC} gives a summary of all selection cuts
imposed to identify RRab stars.

%

\begin{table}
    \caption{Discrimination criteria used to select RRab type stars.}
    \centering
    \label{tab:RRabC}
    \begin{tabular}{@{}lr}
        \hline
        Criteria & Limits \\
        \hline
        Period            & $0.4   - 1.0$ days \\
        $n_{max}$         & $\geq 3$           \\
        Amplitude         & $\leq 2$ mag       \\
        $T_{m}$           & $\geq 0.5$         \\
        \hline                                 \\
    \multicolumn{2}{p{80mm}}{
    Notes: $n_{max}$ is the number of Harmonics used by the AFD, and $T_{m}$
    is the value of the M-statistic as given by equation \ref{eq:CLASS1}.}
    \end{tabular}
\end{table}

\subsection{Photometric Metallicities and Absolute Magnitudes}

Photometric metallicities for the stars in our sample are based on the
relation established in \citet{jjgk96}:

\begin{equation}
{\rm [Fe/H]}_{j}=-5.038-5.394\,P+1.345\,\phi_{31},
\label{eq:CLASS2}
\end{equation}

\noindent where $P$ is the period in days,
$\phi_{31}=\phi_{3}-3\,\phi_{1}$, and the metallicity is given on the
scale of \citet{jj95}. It is important to note that, when computing
the Fourier decomposition coefficients (eq. 2), we require that all
amplitudes are positive and the phases are restricted to the range
between 0 and $2\pi$. We transform the metallicity in equation
\ref{eq:CLASS2} to the standard \citet{zw84} scale using equation 4 of
\citet{jj95}:

\begin{equation}
{\rm [Fe/H]}_{ZW}=\frac{1}{1.431}\left({\rm [Fe/H]}_{j}-0.88 \right),
\label{eq:CLASS3}
\end{equation}

\noindent which in turn can be transformed to the UVES high resolution
spectroscopic scale of \citet{Carretta2009} using:

\begin{equation}
{\rm [Fe/H]}_{UVES}=1.105\,{\rm [Fe/H]}_{ZW}+0.16.
\label{eq:UVES}
\end{equation}

\noindent To check the reliability of these metallicity estimates, we
calculate the $D_{m}$ factor \citep[see][ eq 6]{jjgk96}. Note that at
least 6 orders in Fourier decomposition of the lightcurve are required
to calculate this parameter. Many RRab stars in our sample have 3 or
more orders in their AFD model. Therefore, we produce two $D_{m}$
estimates, one using the available orders only (that we called
$D_{mp}$), and second forcing the AFD to have all 6 orders, as in the
calculations originally made by \citeauthor{jjgk96}. Note that there
are known issues with this metallicity estimate, as the relation tends
to overestimate [Fe/H] at the metal poor end by $\sim 0.3$ dex
\citep[see e.g.][]{jn04,rs05}.

Given the possibility that values based on the \citet{jjgk96}
expression can overestimate the RR Lyrae metallicity at the low end,
it is worth considering possible alternatives to test the robustness
of our results. Recently, a new period-$\phi_{31}$-metallicity
relation has been published by \citet{jnea13}.  The authors use the RR
Lyrae in the {\em Kepler} satellite's field of view to establish the
following transformation:

\begin{equation}
{\rm [Fe/H]} = -8.65-40.12\,P+5.96\,\phi_{31}+6.27\,P\,\phi_{31}-0.72\,\phi_{31}^2,
\label{eq:N13}
\end{equation}
%


\noindent where $P$ is the period of the star in days. Note that this relation
uses the {\em Kepler} $K_p$ system; therefore, it is necessary to transform it
to the standard Johnson $V$ by applying equation 2 of \citet{jnea11}, i.e.
$\phi_{31}(V)=\phi_{31}(K_{p})-0.151(\pm 0.026)$. \citet{jnea13} suggest that
[Fe/H] values obtained in this way are in the UVES scale, but we have checked
that, at intermediate metallicities (as applicable to most of the stars in our
sample), they are also in good agreement with the [Fe/H] values provided by
\citet{jjgk96}, acordingly, we compare them directly, which is the same
approach that \citet{jnea13} uses to compare their results with
\citet{jjgk96}\footnote{Note that Table \ref{tab:RRab} includes the lightcurve
information necessary to derive photometric metallicities in different
scales}. The relation has an intrinsic scatter of only 0.084 dex; however, we
cannot carry out the conventional error propagation as no details of the
covariance matrix are provided by the authors. This makes it impossible to
gauge realistic errors for stars with poorly determined $\phi_{31}$. To this
end, even when we use both scales to study the metallicity distribution in the
halo, we must retain the \citet{jjgk96} metallicity measures for the
estimation of the RRab distances.

Given the metallicity, the absolute magnitude of a RRab star is
calculated according to the following expression, from \citet{mccc08}:
\begin{equation}
M_{V}=(0.23 \pm 0.04){\rm [Fe/H]}_{ZW}+(0.95 \pm 0.13),
\label{eq:CLASS4}
\end{equation}

\noindent which can be converted to obtain absolute magnitudes from the
\citet{Carretta2009} scale using equation \ref{eq:UVES} to obtain:

\begin{equation}
M_{V}=(0.21 \pm 0.04){\rm [Fe/H]}_{UVES}+(0.92 \pm 0.13).
\label{eq:CLASS5}
\end{equation}

\noindent Finally, the unreddened magnitudes are obtained by applying the dust
extinction corrections from the \citet{dsea98} reddening maps. Note that the
color excess, $E(B-V)$, might be overestimated for values greater than 0.15
mag \citep{ArceGoodman1999}. These values might be corrected by using the
relation given in equation 1 of \citet{Bonifacio2000}. Note, however, that
these corrections are small ($\sim10\%$) for $E(B-V)<0.3$. Since our stars are
typically located in zones of high Galactic latitude, where the extinction is
low, we do not take into account this effect. Nevertheless, we include the
values of $E(B-V)$ with the detected RRab's. 
lower than 0.3


\subsection{Completeness and Reliability}\label{sssec:COMP}

\begin{figure}
    \includegraphics[width=84mm]{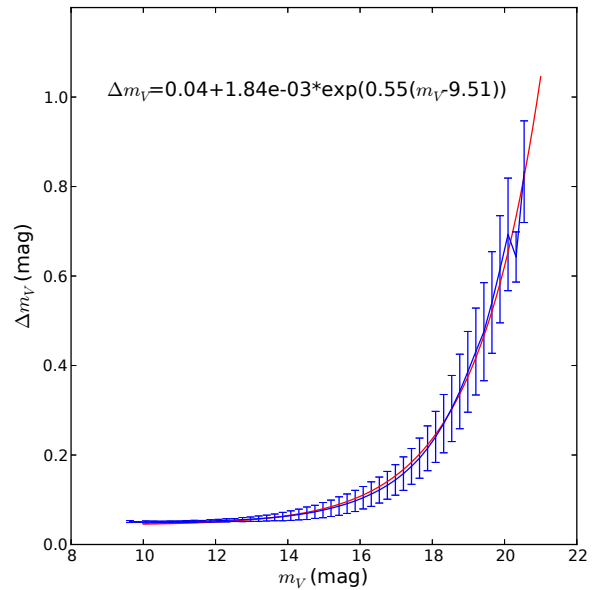}
    \caption{Photometric error as a function of apparent magnitude for
      the SSS survey. Red line represents the exponential model fit to
      the mean error in each magnitude bin. This is the model used to
      produce synthetic lightcurves as described in Section
      \ref{sssec:COMP}.}
    \label{fig:fig10}
\end{figure}

To check the reliability of our algorithms, we generate synthetic
lightcurves based on \citet{cl98} RRab templates. To make the
lightcurves as similar as possible to the ones observed by SSS, we
studied how the error of the observations change with
magnitude. Figure \ref{fig:fig10} shows the behaviour of the error
with increasing apparent magnitude. To model the error evolution as a
function of magnitude, an exponential model is used:
\begin{equation}
\sigma_{M}(M) = 0.04 + 1.84\times10^{-3} \exp(0.55[M-M_{min}]),
\label{eq:RES1}
\end{equation}
%


\begin{figure}
    \includegraphics[width=84mm]{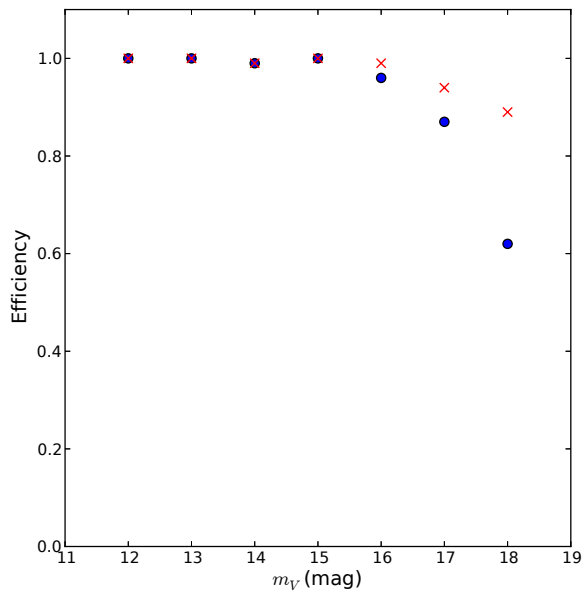}
    \caption{Reliability of period (red crosses) and RRab (filled blue
      circles) selection as a function of magnitude. Each datapoint
      gives the fraction of stars recovered from the synthetic sample
      as a function of the apparent magnitude.}
    \label{fig:fig11}
\end{figure}

\noindent where $M_{min}=9.51$ corresponds to the brightest star in our
sample.  To generate the synthetic RRab, pulse amplitudes as well as phases of
random stars in the sample are used; the errors are obtained from equation
\ref{eq:RES1}, and the noise is drawn from a normal distribution with sigma as
given by equation \ref{eq:RES1}. For each magnitude bin, 100 synthetic RRab
are generated, in the range $12 <V <18$.  We then run the APS algorithm and
count the number of mock RRab we can recover as a function of apparent
magnitude. Figure~\ref{fig:fig11} illustrates that at least 60\% of the RRab
with $V=18$ are recovered. The same figure also reveals that the efficiency of
the period estimation is even higher, reaching $\sim 90\%$. It is important to
mention that the APS was set to have the cleanest sample possible in a fully
automatic mode which explains why the efficiency drops down at the faint end.
This is, of course, unavoidable since at these magnitudes the photometric
error becomes comparable to the pulse amplitudes of RRab stars, making it
difficult for a fully automated procedure to identify correctly the exact type
of variability. Nevertheless, as part of the output of the program, the
information on the rejected stars is recorded, so that the faint RRab
discarded by our search algorithm can be recovered in the future.

The analysis described above concerns only the efficiency of the period
determination and the RRab classification. The actual completeness of our
sample, i.e. the amount of stars expected to be recovered versus the actual
number of stars measured, needs to take into account the variability and the
periodicity checks described in Section \ref{sec:ODC}. An example of an
empirical approach to check the completeness of the RRab sample can be found
in \citet{adea13a}, which reports an efficiency of $\sim70\%$ for the RRab
sample found by the CSS. Since both the CSS and the SSS performed the same
steps to classify RRab, this could serve as an adequate estimate of the
overall completeness of the SSS RRab sample presented in this paper. In fact,
we have verified this statement by counting all stars that lie in the SSS
footprint and are recorded as RRab in the VSX catalog \citep{Watson2006}.
There are 1919 RRab in the VSX of which we identify 1321, resulting in a
completeness level of $\sim 69 \%$. If we now remove detections along the
borders of the survey, where objects can be spuriously included even if they
were not observed sufficiently often by the SSS, we identify 1075 out of 1521
stars, raising the completeness to $\sim 71 \%$. These tests illustrate that
our efficiency is a strong function of the temporal sampling, increasing in
the areas of the sky visited by the SSS more frequently. For example, if we
further limit the region of interest to $190 < \alpha (\rm{deg}) < 235$ and
$-23 < \delta (\rm{deg}) < -35$, i.e. the area sampled the densest by the SSS,
we detect $84 \%$ of the RRab's in the VSX database. While this is
encouraging, completeness estimates based on a comparison between our
detections and the VSX entries should be taken with due caution, as the VSX
catalog is highly inhomogeneous and has a much brighter cutoff with respect to
SSS. Most importantly, we note that the VSX catalog is highly incomplete  in
the Southern hemisphere, as it contains less than 10\% of the total number of
RRab's we have discovered.


%


\begin{figure*}
    \includegraphics[width=168mm]{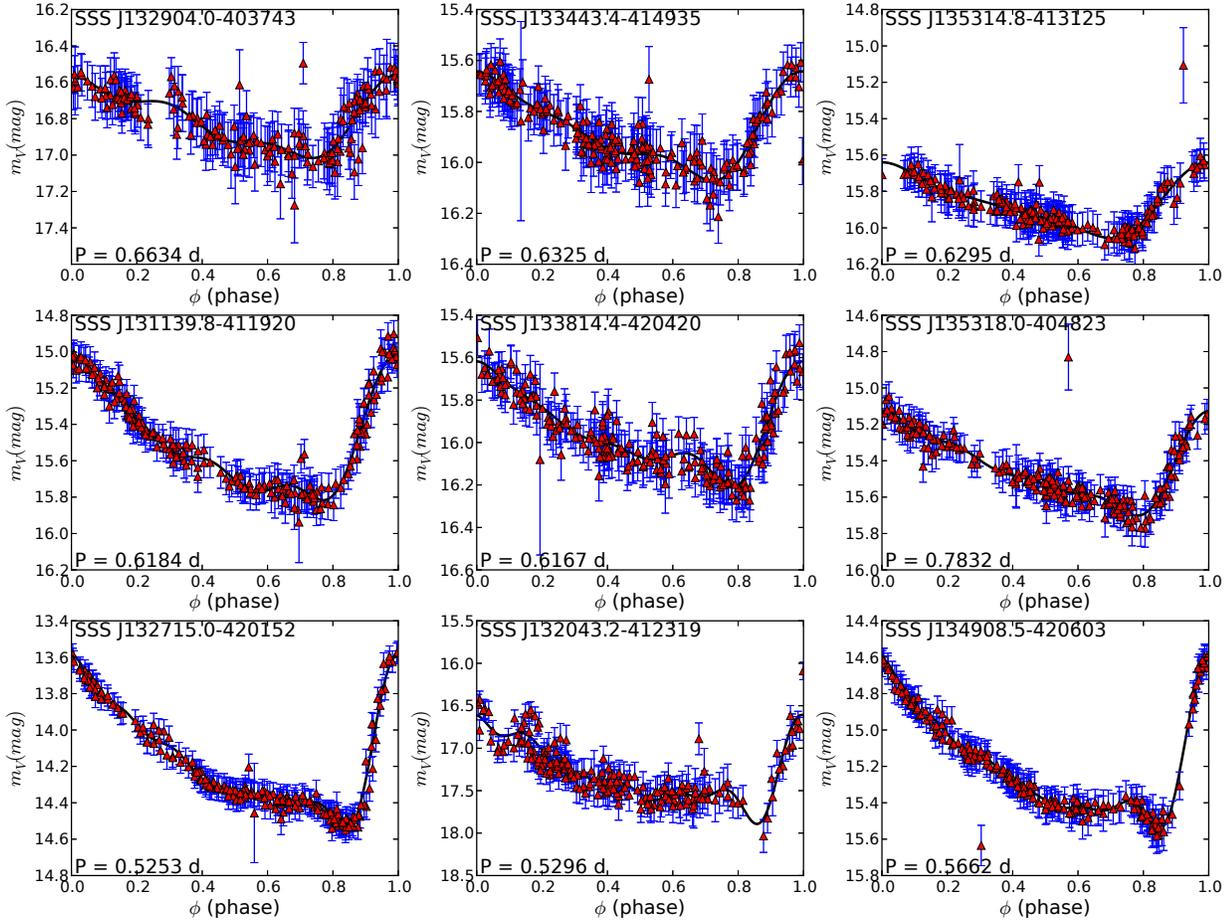}
    \caption{Examples of RRab lightcurves. From top to bottom,
      lightcurves of sample RRab variables that require 3 (top), 4
      (middle), and 6 (bottom) Fourier harmonics in the model fit. The ID and 
      the period in days are given for each star.}
    \label{fig:fig7}
\end{figure*}

\section{Galactic stellar halo with new RR Lyrae}\label{ssec:RES}

\begin{figure*}
    \includegraphics[width=168mm]{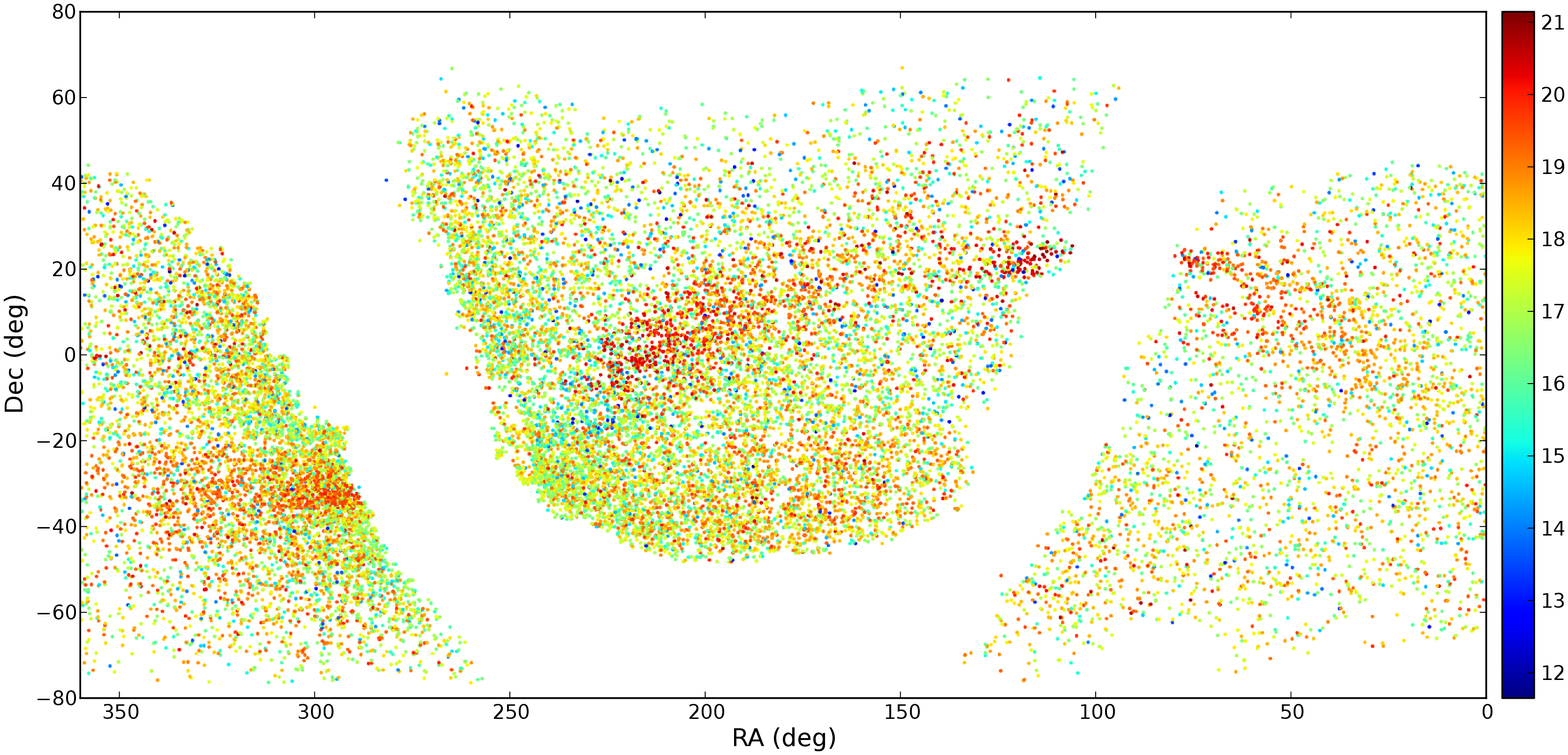}
    \caption{Distribution of all RRab stars discovered in the three
      Catalina surveys in equatorial coordinates. The color represents
      the magnitude of each star, following the color code on the right. The obvious stream-like pattern
      running across the Figure is the Sagittarius tidal tail
      system. The new RRab found in the SSS lie below
      $\rm{Dec}=-20^{\circ}$. These trace the beginning of the trailing arm
      starting at $\rm{RA}\sim290^{\circ}$ and $\rm{Dec}\sim-30^{\circ}$.}
    \label{fig:fig7a}
\end{figure*}

From the original sample of $18,\!288$ variable star candidates with periods
between 0.3 and 4 days, we have selected $10,\!540$ RRab stars, of which
$8,\!869$ correspond to new discoveries, 231 correspond to stars previously
reported as RRab by \citet{adea13a}, and the remaining $1,\!440$ correspond to
stars previously available in the literature. Figure~\ref{fig:fig7} displays
several examples of light curves for the RR Lyrae identified in our search.
Figure \ref{fig:fig7a} presents the distribution of RRab on the sky, using
discoveries from all three Catalina surveys. For all stars in the SSS we
provide metallicity, distance, Galactic position, and lightcurve shape
information (through the AFD parameters). Table \ref{tab:RRab} shows an
extract from our RRab catalog.


\begin{figure*}
    \includegraphics[width=168mm]{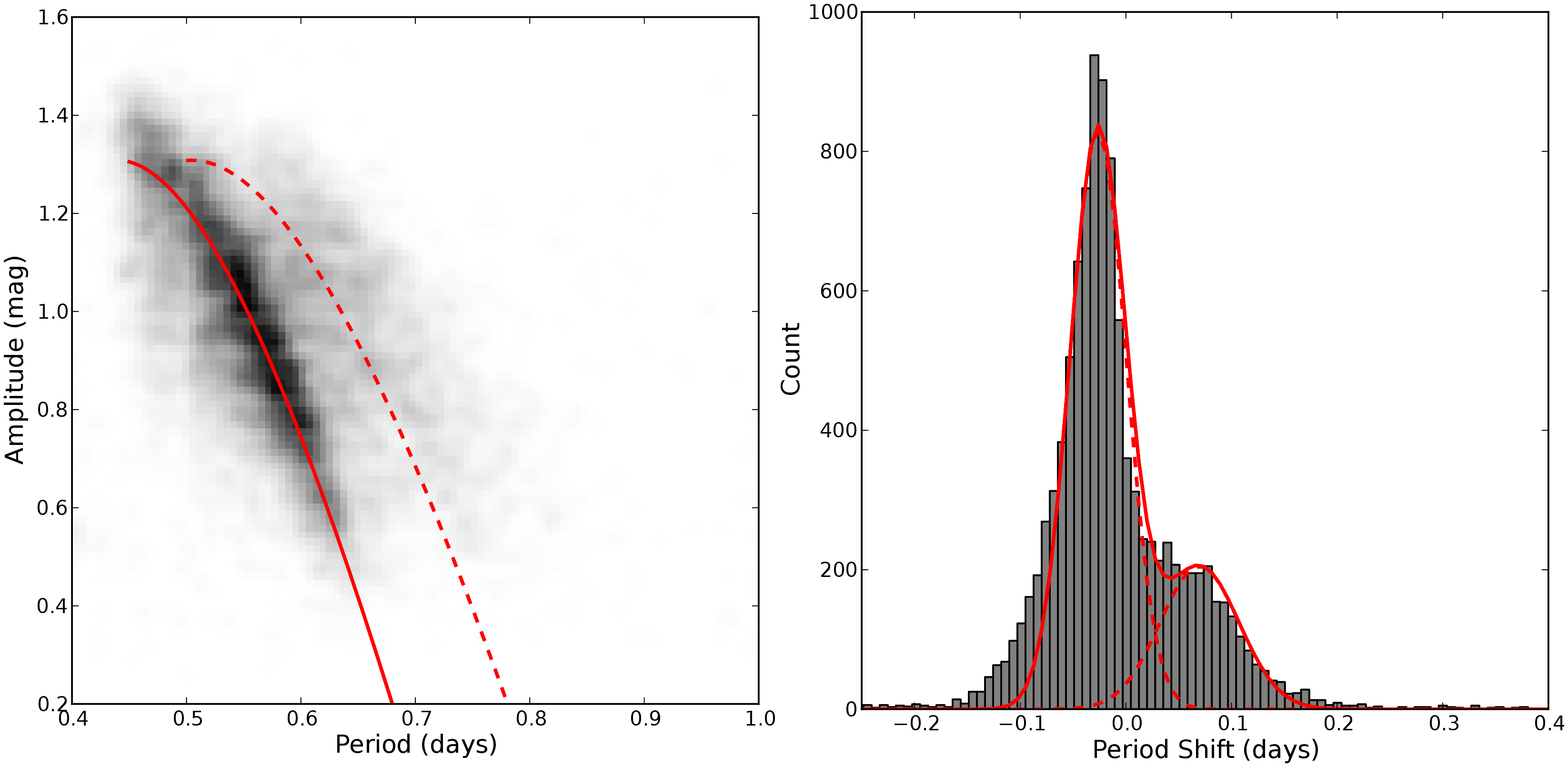}
    \caption{{\it Left:} Period-amplitude diagram of the selected RRab
      stars. For reference, lines corresponding to the OoI (solid) and
      OoII (dashed) objects based on Equation 11 of \citet{mzea10} are
      overlayed. {\it Right:} Histogram of the period difference (at
      fixed amplitude) between the detected RRab's and the OoI
      reference line. The red lines shows a two-Gaussian fit to the
      histogram. This distribution is similar to what is shown in
      \citet{adea13a} for the RRab identified in the Northern
      hemisphere. Similarly, our result suggests an overall bi-modal
      population composed by both OoI and OoII RRab stars, with the
      OoI sub-sample being notably larger compared to the OoII
      one. The excess of stars observed in low-period-shift regime may 
      be caused by RR Lyraes that present the Blazhko effect.}
    \label{fig:fig7b}
\end{figure*}

Figure~\ref{fig:fig7b} presents the distribution of the discovered
RRab stars across the period-amplitude diagram (also known as Bailey
diagram). As shown previously by \citet{adea13a}, amplitudes of RRab
stars in our sample are systematically reduced by 0.15 mag when
compared to the typical RRab amplitudes. As a result, we provide a
correction to the amplitudes to account for this effect. The diagram
shows the \citet{mzea10} Oosterhoff type-I (OoI) and Oosterhoff
type-II (OoII) lines, and the period shift with respect to the OoI
line is shown on the right panel. This suggests that our data is
composed of stars belonging to both OoI and OoII types, with the OoI
population clearly being the dominant one. As in Drake et al. (2013a),
we too find that the distribution is bimodal, with relatively few
stars present between the canonical OoI and OoII loci.

\begin{table*}
    \scriptsize
    \begin{minipage}{180mm}
    \caption{RRab sample selection}
    \label{tab:RRab}
    \begin{tabular}{@{}crrrrrrrrrcrr}
        \hline
        ID                   &RA       &Dec      &$m_{V}$&Period &Amp  &$n_{obs}$&$d_{H}$& [Fe/H]&E(B-V)&$D_{m}$&$\phi_{31}$&ID$_{alt}$           \\ 
                             &(deg)    &(deg)    &(mag)  &(days) &(mag)&         &(Kpc)  & (dex) &      &       &(rad)      &                     \\   
        \hline
        SSS\_J055136.9-143214&87.90386 &-14.5371 &12.87  &0.51901&0.77 &82       &2.24   &-1.38  &0.176 &5.3    &4.97       &BF\_Lep              \\
        SSS\_J085855.1-151229&134.72958&-15.20811&15.8   &0.58286&0.75 &95       &10.32  &-1.87  &0.047 &42.7   &4.29       &...                  \\
        SSS\_J090949.5-153559&137.45624&-15.59973&12.02  &0.50775&1.05 &119      &1.75   &-1.16  &0.082 &5.6    &5.38       &XX\_Hya              \\
        SSS\_J091750.5-151433&139.46041&-15.24257&12.9   &0.66385&0.38 &124      &3.05   &-2.59  &0.055 &85.8   &5.47       &VSX\_J091750.5-151433\\
        SSS\_J092942.0-154231&142.42515&-15.70864&15.61  &0.49154&1.14 &126      &8.48   &-1.46  &0.0686&5.5    &4.99       &...                  \\
        SSS\_J093650.9-150241&144.21208&-15.04463&17.57  &0.52288&1.06 &115      &23.52  &-1.46  &0.069 &12.7   &6.18       &...                  \\
        SSS\_J093543.0-150150&143.92933&-15.03052&15.1   &0.64336&0.26 &115      &6.96   &-1.27  &0.076 &131.7  &5.49       &...                  \\
        SSS\_J094138.1-155627&145.40875&-15.94082&16.54  &0.50207&1.08 &117      &9.58   &-1.78  &0.0681&49.7   &5.5        &CSS\_J094138.1-155626\\
        SSS\_J095250.9-143253&148.21227&-14.54792&16.31  &0.5016 &1.19 &118      &12.46  &-1.31  &0.0574&4.5    &4.5        &CSS\_J095251.0-143253\\
        SSS\_J094959.6-141035&147.4988 &-14.17705&15.89  &0.57123&0.77 &114      &10.85  &-1.56  &0.052 &6.9    &4.66       &CSS\_J094959.6-141035\\
        SSS\_J100008.3-153936&150.0344 &-15.66008&17.08  &0.49428&1.15 &126      &16.16  &-1.54  &0.0506&12.3   &5.66       &...                  \\
        \hline
    \end{tabular}
    Notes: Column 1 gives the SSS id; Column 2 and 3 give the right ascension and declination; Column 4 gives the average magnitude, as given by the AFD; Column 5 gives the period; Column 6 gives the amplitude of the lightcurve, as given by the AFD; Column 7 gives the number of photometric observations; Column 8 gives the distance to the Sun, in kpc; Column 9 gives the metallicity, as given by eq.~\ref{eq:CLASS2} and transformed to the UVES scales with eq.~\ref{eq:UVES}; Column 10 gives the reddening, as given by \citet{dsea98}; Column 11 give the value of $D_{m}$, which gives us sn estimate of the reliability of the photometric metallicities according to \citet{jjgk96}; Column 12 gives the value of $\phi_{31}$; Column 13 gives the id for sources that were previously known.

    This table is available in its entirety in a machine-readable form in the online journal. A portion is shown here for guidance regarding its form and content.
    \end{minipage}
\end{table*}

\subsection{Halo Metallicity}\label{ssec:MHM}

\begin{figure}
    \includegraphics[width=82mm]{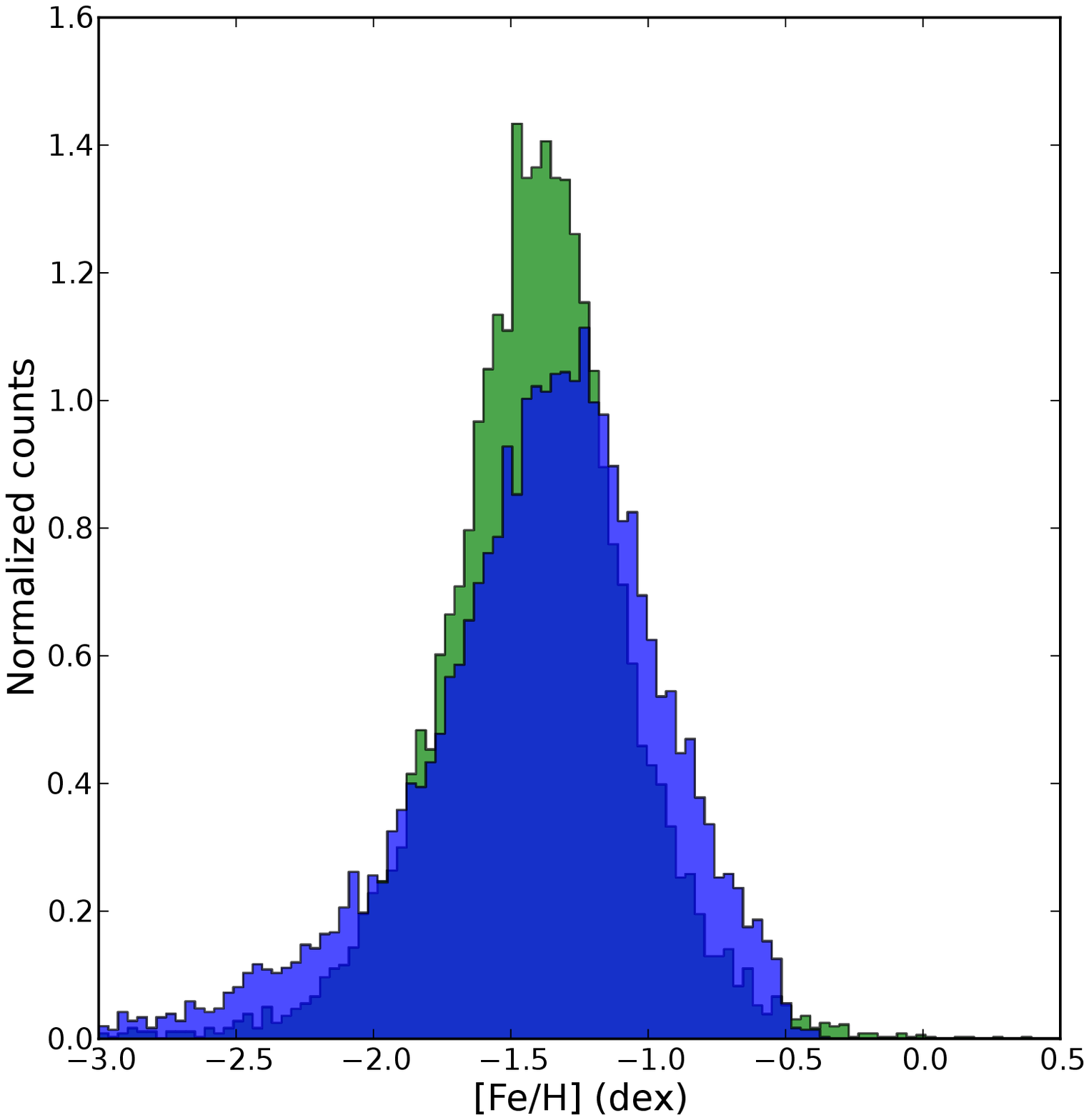}
    \caption{Metallicity distribution of the SSS RRab stars according
      to \citet{jjgk96} (green) and \citet{jnea13} (blue)
      relations. The histograms are normalized to the same area. The 
      green histogram can be described by a Gaussian
      distribution with a peak at ${\rm [Fe/H]}=-1.40$~dex and a
      dispersion of $0.20$~dex, whereas the blue histogram shows the
      presence of a metal-poor tail and an excess of stars in the
      range $-1.0 < {\rm [Fe/H]} <-0.5$. The existence of these
      differences in the metallicity estimated based on the RRab
      lightcurve shape is in agreement with previous studies
      highlighting that \citet{jjgk96} metallicities are typically
      higher than expected in the metal-poor regime. Also note that a new
      discrepancy appears in the metal-rich regime, where the \citet{jjgk96}
      relation seems to underestimate the ${\rm [Fe/H]}$ values.}
    \label{fig:fig12}
\end{figure}

\citet{Carretta2009} presents a new homogeneous metallicity scale, which is
calibrated using high resolution spectra from UVES. It also provides
transformations from the most used metallicity scales. The fact that this
scale has been calibrated using a large sample of GC red giants with
homogeneously measured metal abundances from high resolution spectroscopy,
makes this scale a good candidate to homogenize the metallicity measurements
for RR Lyrae. In this sense, we transform all our metallicities to the UVES
metallicity scale provided by \citet{Carretta2009}. The total number of RRab
stars with metallicities in our sample is $10,\!540$. This number is reduced
slightly after the removal of obvious outliers, that is, the stars whose
metallicities fall outside the range $-3.0<{\rm [Fe/H]}<0.5$. This leaves a
total of $10,\!369$ stars with photometric metallicities, an ample resource to
probe the properties of the Galactic halo out to 40 kpc.
Figure~\ref{fig:fig12} shows two metallicity distributions, one found using
the \citet{jjgk96} relation (green) and the results of the alternative ${\rm
[Fe/H]}$ determination with formulae from \citet{jnea13} (blue). As the error
propagation cannot presently be implemented for the latter, in order to
compare the two distributions, we have to assign uncertainties found using the
\citet{jjgk96} relation to ${\rm [Fe/H]}$ values obtained according to
\citet{jnea13}. For the \citet{jjgk96} metallicities, the histogram has a
shape that can be well approximated by a Gaussian with a mean metallicity of
${\rm   [Fe/H]}=-1.41$~dex and a standard deviation of $0.27$~dex. Note that
if we limit the sample to the stars with reliable metallicity estimates in
accordance with the criteria from \citeauthor{jjgk96} (i.e., with values of
$D_{m}<5$; this gives a total of $1,\!880$ stars), the mean of the
distribution remains the same at ${\rm [Fe/H]}=-1.40$~dex, but the dispersion
shrinks slightly, to $0.20$~dex. This result appears to be in good agreement
with independent, spectroscopic analyses of the halo RR Lyrae and blue
horizontal-branch stars \citep[e.g.,][and references therein]{tkea00,adea13a}.
Small differences, such as the presence of a metal-rich tail and/or lack of
an extremely metal-poor tail (${\rm   [Fe/H]}\ll -2$~dex), could be due to
small systematic differences in the metallicity scale adopted, as well as the
lack of very metal-poor calibrators in the \citet{jjgk96} sample. Naturally,
the RR Lyrae metallicity distribution cannot be directly compared with the one
derived using other stellar tracers, be it main-sequence stars
\citep[e.g.,][and references therein]{daea12} or giants. This is because the
formation mechanism of RR Lyrae stars is far from uniform as a function of
metallicity: it peaks around ${\rm[Fe/H]}\approx-1.5$
\citep[e.g.,][]{mcpd93,al95,tkea09}. Note that the distribution of the
\citet{jnea13} metallicities mostly agrees with that built using the
\citet{jjgk96} prescription, except for the appearance of a metal-poor tail,
confirming that the \citet{jjgk96} method does indeed overestimate the
metallicities at the very low end. Interestingly, we also find an excess of
stars in the metal-rich end, where it seems that the \citet{jjgk96} method is
underestimating the ${\rm[Fe/H]}$ values.

\begin{figure}
    \includegraphics[width=0.49\textwidth]{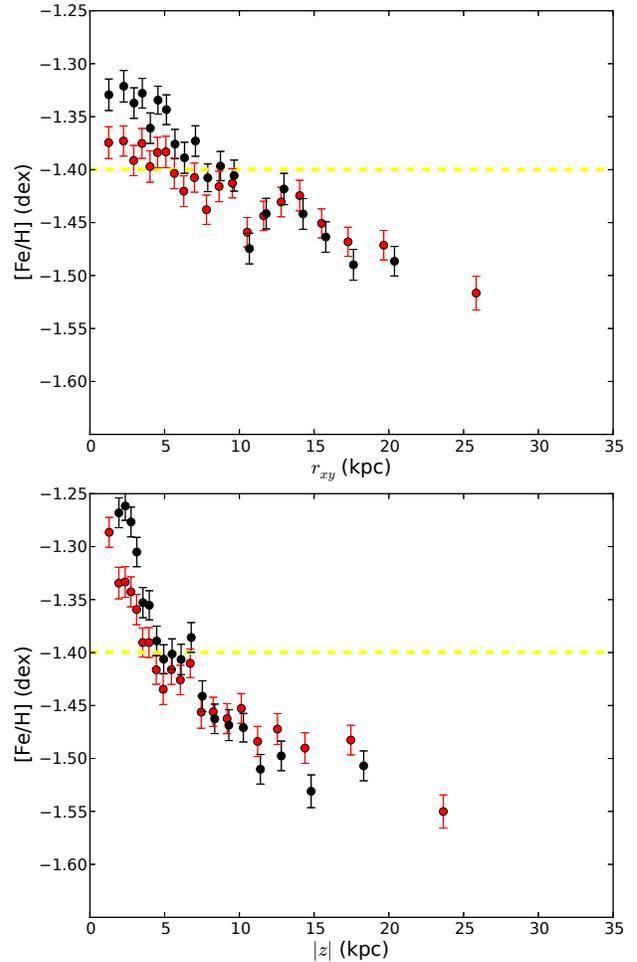}
    \caption{Spatial variations of the RRab metallicity in the SSS. In
      both panels, the bin-size along the x-axis is varied to make
      sure that there are 500 stars in each bin. Red (black) symbols
      show average RRab metallicities obtained with the \citet{jjgk96}
      (\citet{jnea13}) relation. {\it Top:} Mean metallicity (filled
      circles) and the associated error for bins of Galactocentric
      cylindrical radius. Note a constant metallicity in the first 4
      to 5 kpc from the Galactic centre. This is followed by a modest
      but statistically significant negative metallicity gradient for
      RRab's with $7 < r_{xy} (\rm{kpc}) < 25$. {\it Bottom:} Mean metallicity
      (filled circles) and the associated error for bins of distance
      above Galactic plane $|z|$. Note a steep ${\rm [Fe/H]}$ gradient
      in the first 5 kpc, followed by a much shallower slope for stars
      with $7 < |z| (\rm{kpc}) < 25$. It can also be seen that the
      \citet{jnea13} relation indicates a higher concentration of more
      metal-rich stars near the disk.}
    \label{fig:fig13}
\end{figure}

Let us study the spatial variations of the RRab metallicity in the
halo. For the analysis described below, to obtain reliable space
coverage, we choose to include stars with $D_{m}>5$ as their
metallicity distribution does not look significantly different from
that of the stars with reliable ${\rm [Fe/H]}$.
Figure~\ref{fig:fig13} shows the mean RRab metallicity as a function
of Galactic cylindrical radius, $r_{xy}=\sqrt{x^{2}+y^{2}}$ (top
panel), and as a function of the distance from the Galactic plane $z$
(bottom panel). The bin size for each of the x axes is chosen
adaptively so that there are 500 stars in each bin. The figure gives
the mean and its error for each bin, for both the \citet{jnea13}
metallicities (black), as well as the \citet{jjgk96} metallicities
(red). Note that while the ${\rm [Fe/H]}$ spread in each bin is large
($\sigma\sim0.35$~dex), the uncertainty of the mean, due to moderate
uncertainties on individual metallicities and large number of stars in
our sample, remains small ($\sim0.02$~dex).

The top panel of Figure \ref{fig:fig13} reveals a small but
nonetheless very clear negative metallicity gradient both in
Galactocentric distance and in distance from the Galactic plane,
regardless of the ${\rm [Fe/H]}$ recipe used. Note that as a function
of Galactocentric radius, the mean metallicity starts to drop at
$\sim7$~kpc, having remained approximately constant in the first few
kpc. This is in good agreement with previous studies of the
metallicity gradient \citep[e.g.,][]{dsea09,kkea06}, where the authors
claim a lack of evidence for a metallicity gradient in the first
$\sim10$~kpc.  However, as Figure~\ref{fig:fig13} makes apparent, this
behaviour changes at larger distances from the Galactic center. Out to
$r_{xy}\sim10$ kpc, the \citet{jnea13} estimate of metallicity is
consistently higher (by approximately one standard deviation) as
compared to that of \citet{jjgk96}, which suggests that the excess of
stars with ${\rm [Fe/H]}>-1$ according to \citet{jnea13} (as seen in
Figure~\ref{fig:fig12}) is preferentially located within the Solar
radius. Moreover, as the lower panel illustrates, these stars live
very close to the Galactic disk. This panel also reveals an obvious
dependence of the mean RRab metallicity on $|z|$ as well as a
concentration of high metallicity RRab in the first few kpc, which is
consistent with the hypothesis that the high metallicity RRab's belong
primarily to the Milky Way's disk \citep{kkea06}. We also note the
possible detection of a small bump in the radial profile of the mean
RRab metallicity at $r_{xy}\sim13$ kpc. We have checked the possibility that
the bump is due to presence of the Sagittarius stream in our
dataset. However, the removal of the likely Sgr members has no
noticeable affect on this feature.

\subsection{A sweep for obvious halo sub-structure}\label{ssec:MHS}

In this section we carry out a simple overdensity search using the
discovered RRab as tracers. Our method takes advantage of the
available 3D information and picks up the most obvious (largely,
previously unknown) overdensity candidates in the southern celestial
hemisphere. To find significant enhancements in RRab density, we
calculate the local stellar density around every star and compare it
to the prediction of a smooth stellar halo model. We keep track of
those positions where the observed number of stars is substantially
higher than the expected number, as given by the model.The coordinates
as well as some of the structural properties of the detected
candidates are given in Table~\ref{tab:ovdc}.

The observed density is calculated using the N-th neighbor method,
originally proposed by \citet{ad80}. At each trial position, this
algorithm takes the distance to the N-th nearest star as the radius of
the sphere containing N stars. Then the number density at the selected
position is simply given by:

\begin{equation}
\rho_{l}=\frac{N}{\frac{4}{3}\pi d_{N}^{3}},
\label{eq:SSTRUCT1}
\end{equation}

\noindent where $d_{N}$ is the distance to the N-th star from the
selected position. The local density estimate is then compared to the
prediction of the halo model, in particular, we use the model proposed
by \citet{bsea10}. This is a flattened power-law as defined by:

\begin{equation}
\rho_{m}=\rho_{0}\left[ \frac{R_{\odot}}{\sqrt{x^{2}+y^{2}+\left(^{z}/_{q_{H}} \right)^{2}}} \right]^{n_{H}},
\label{eq:SSTRUCT2}
\end{equation}

\noindent where $\rho_{0} =4.2\,\mbox{kpc}^{-3}$ is the number density
of RRab stars at $R_{\odot}=8.0$ kpc \citep{kvrz06}, $q_{H} =0.64$
stipulates the oblateness of the halo \citep{bsea10}, $n_{H}=2.77$ is
the power-law index \citep{mjea08}, and $x$, $y$, and $z$ are
Cartesian coordinates in the Galacto-centric system. In order to
select the stars associated with each overdensity candidate, we
determine the extent of the sub-structure by calculating 3D density
contours around its peak. The outermost contour is set to a density
$0.5$ times smaller than the peak density of the candidate. The
significance is then obtained by assuming the Poissonian statistics
for the incidence of the sub-structure member stars.

\begin{figure*}
    \includegraphics[width=168mm]{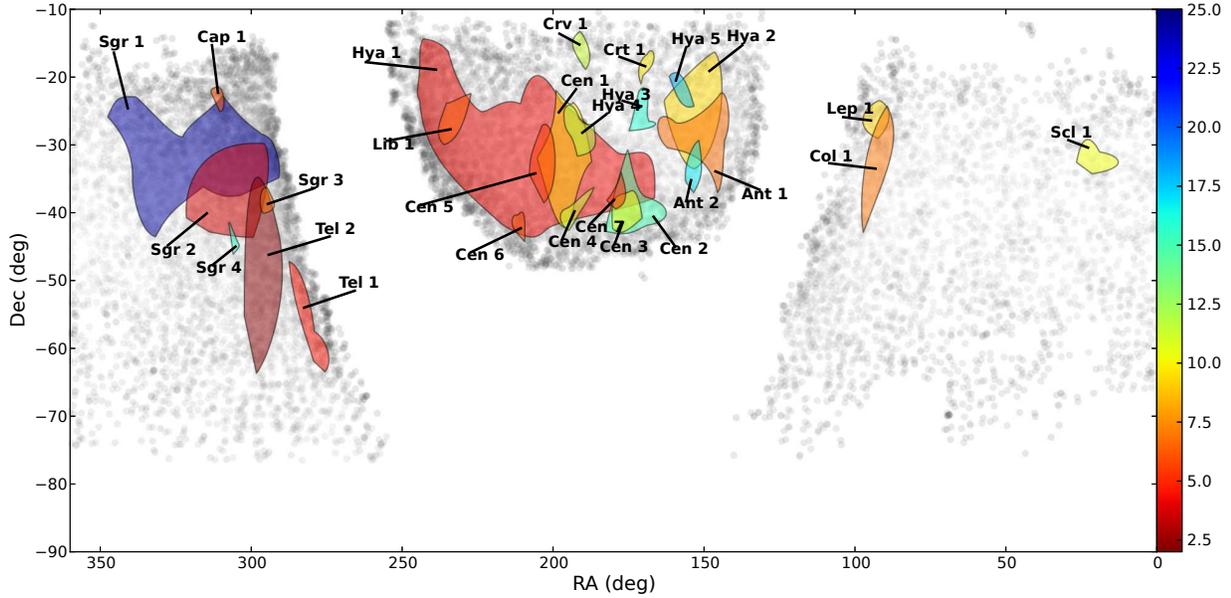}
    \caption{Locations and shapes of the SSS RRab overdensity
      candidates as presented in Table \ref{tab:ovdc}. The shapes
      are labeled according to the reference ID given in Column 1 of
      the table, and the color of each shape reflects the mean
      heliocentric distance to the constituent stars (see color-bar
      for reference in kpc). The background greyscale distribution
      shows all SSS variable candidates; the shades of grey are coded according
      to the logarithm of the number of observations.}
    \label{fig:fig13a}
\end{figure*}

By applying the method described above to our sample of SSS RR Lyrae, we have
discovered several interesting candidates; in particular, there are 12
possible objects each with significance greater than $3~\sigma$. Table
\ref{tab:ovdc} summarizes the properties of all candidates with significance
greater than $2~\sigma$. This table includes information regarding the shape
of the candidate by listing its extent in RA and Dec, as well as its
heliocentric distance. To illustrate the distribution of the sub-structure
candidates in the Galaxy, we have plotted their approximate shapes in
equatorial coordinates in Figure~\ref{fig:fig13a}. Here, the color of each
polygon represents the mean heliocentric distance of the stars bounded by the
region. All SSS variable candidates are plotted in greyscale in the background
of the figure, to show the survey's limits. Please bear in mind that the
objects that are located closer will naturally appear bigger in this figure.
Additionally, it is also worth noting that the second most prominent
candidate, Hya\,1, situated very close to the Galactic center, has a very
large size. This may simply be caused by the algorithm selecting too big an
area around a genuine overdensity close to the MW center (note that the
significance, in this specific case, is $\sim 9~\sigma$). This could be due to
the model (see equation \ref{eq:SSTRUCT2}) producing an underestimate of the
true number of RRab stars in this region of the Galaxy. Note that the same may
apply to Col\,1 and Tel\,1 which are close to the MW disk. In their case, its
significance might also be inflated due to a poor estimation of the amount of
background stars by the model. Conversely, some of the candidates in the range
$130\deg \lesssim {\rm RA} \lesssim 250\deg$  could plausibly be related to
the VOD.  To confirm the nature of these overdensity  candidates, deep wide-
area imaging and spectroscopic data would be necessary.

\begin{table*}
    \scriptsize
    \begin{minipage}{150mm}
    \caption{Overdensity Candidates in the Galactic Halo}
    \label{tab:ovdc}
    \begin{tabular}{@{}lrrrrrrrrrrr}
        \hline
             ID (1)&ID (2)&S         &RA   &Dec  &$n_{RRL}$&$m_{V}$&$m_{V,\rm{min}}$&$m_{V,\rm{max}}$&$\Delta D$&$\Delta\,\alpha\,\cos{\delta}$&$\Delta\,\delta$  \\ 
                   &      &($\sigma$)&(deg)&(deg)&         &(mag)  &(mag)           &(mag)           &(kpc)     &(deg)                         &(deg)             \\
        \hline
        Sgr\,1&SSSOc\_J203007-320757&15.79&307.53&-32.13&327&17.71&16.71&18.38&13.0&48.6&20.8\\
        Hya\,1&SSSOc\_J145754-283033&9.10 &224.48&-28.52&179&14.03&12.56&15.22&4.2 &70.9&29.6\\
        Cen\,1&SSSOc\_J130248-320118&4.74 &195.70&-32.02&46 &15.63&15.30&16.05&2.4 &15.3&20.5\\
        Cen\,2&SSSOc\_J112907-394650&4.41 &172.28&-39.78&35 &16.72&16.32&17.12&4.4 &17.8&13.0\\
        Hya\,2&SSSOc\_J100540-245134&4.15 &151.42&-24.86&29 &15.61&15.27&15.97&2.7 &19.0&17.9\\
        Sgr\,2&SSSOc\_J202727-372132&4.06 &306.87&-37.36&36 &13.67&12.83&14.18&2.0 &23.0&14.6\\
        Hya\,3&SSSOc\_J111948-252007&3.84 &169.95&-25.34&21 &16.75&16.53&17.11&3.8 &9.6 &7.2 \\
        Hya\,4&SSSOc\_J124106-281257&3.84 &190.28&-28.22&27 &16.18&15.98&16.61&2.9 &11.3&8.3 \\
        Tel\,1&SSSOc\_J184232-553460&3.46 &280.64&-55.58&21 &14.09&13.73&14.48&1.1 &8.0 &16.2\\
        Cen\,3&SSSOc\_J114153-393448&3.33 &175.47&-39.58&20 &15.92&15.60&16.37&3.3 &8.3 &7.7 \\
        Cen\,4&SSSOc\_J124936-391616&3.20 &192.40&-39.27&16 &16.12&15.98&16.31&1.3 &9.4 &6.3 \\
        Ant\,1&SSSOc\_J101002-282309&3.03 &152.51&-28.39&24 &14.95&14.68&15.22&2.1 &23.5&15.3\\
        Crv\,1&SSSOc\_J124414-155910&2.99 &191.06&-15.99&15 &16.00&15.71&16.31&2.6 &9.1 &5.6 \\
        Tel\,2&SSSOc\_J194146-495134&2.82 &295.44&-49.86&17 &12.39&11.87&12.95&0.9 &10.0&34.1\\
        Sgr\,3&SSSOc\_J193812-382955&2.76 &294.55&-38.50&17 &15.10&14.77&15.34&0.8 &6.3 &4.4 \\
        Cen\,5&SSSOc\_J133306-331937&2.75 &203.28&-33.33&18 &14.41&14.21&14.66&0.8 &9.1 &13.0\\
        Lib\,1&SSSOc\_J153446-265727&2.75 &233.69&-26.95&22 &15.01&14.62&15.38&1.2 &11.3&7.5 \\
        Ant\,2&SSSOc\_J101614-331524&2.56 &154.06&-33.26&13 &16.94&16.76&17.05&1.1 &5.7 &7.7 \\
        Cen\,1&SSSOc\_J140450-143546&2.50 &211.21&-41.60&13 &14.78&14.57&15.03&0.7 &5.1 &5.7 \\
        Hya\,5&SSSOc\_J103240-212215&2.50 &158.17&-21.37&11 &17.00&16.81&17.16&1.9 &7.6 &7.0 \\
        Cap\,1&SSSOc\_J204404-230214&2.46 &311.02&-23.04&10 &14.68&14.59&14.80&0.5 &5.3 &3.7 \\
        Cen\,7&SSSOc\_J115216-365325&2.45 &178.07&-36.89&7  &14.28&14.09&14.42&0.4 &6.7 &5.4 \\
        Lep\,1&SSSOc\_J061057-254429&2.44 &92.74 &-25.74&9  &15.59&15.41&15.79&1.5 &13.2&5.6 \\
        Col\,1&SSSOc\_J060924-321527&2.37 &92.35 &-32.26&12 &14.89&14.44&15.39&2.5 &12.3&18.6\\
        Sgr\,4&SSSOc\_J202138-435633&2.36 &305.41&-43.94&9  &16.72&16.60&16.89&1.6 &2.9 &5.4 \\
        Scl\,1&SSSOc\_J011639-313912&2.18 &19.16 &-31.65&9  &15.72&15.46&15.95&1.9 &15.0&6.7 \\
        Crt\,1&SSSOc\_J111824-181426&2.17 &169.60&-18.24&10 &15.64&15.44&15.81&1.1 &8.2 &6.0 \\
        \hline
    \end{tabular}
    Notes: Column 1 gives a reference id; Column 2 gives the
    Overdensity id; Column 3 gives the significance in standard deviations; Columns
    4 and 5 give the mean right ascension and declination of the
    stars of the overdensity; Column 6 gives the number of stars
    associated with the overdensity; Column 7 gives the mean magnitude of
    the stars of the overdensity; Columns 8 and 9 give the minimum and
    maximum magnitude of the stars of the overdensity, respectively; Column 10 gives
    the overdensity's extension along the line of sight, in kpc; Columns 11 and 12
    give the extension in equatorial coordinates.
    \end{minipage}
\end{table*}

\subsection{Sagittarius stream in SSS data}

The most prominent of the sub-structure candidates detected by the
algorithm described above has a striking significance of $>15~\sigma$,
and an impressive total of 327 RRab stars associated with it. The mean
position of the constituent stars is at $\alpha = 307.53^{\circ}$,
$\delta = -32.132^{\circ}$, at a distance of $\sim 24$ kpc from the
Sun and $\sim 18$ kpc from the Galactic center. The overdensity has an
elongated, stream-like shape, with a length of about 17 kpc and a
width of about 6 kpc. Given its 3D position, it is almost certainly a
part of the Sagittarius stream. Therefore, its designated name is
Sgr\,1.

\begin{figure*}
    \includegraphics[width=168mm]{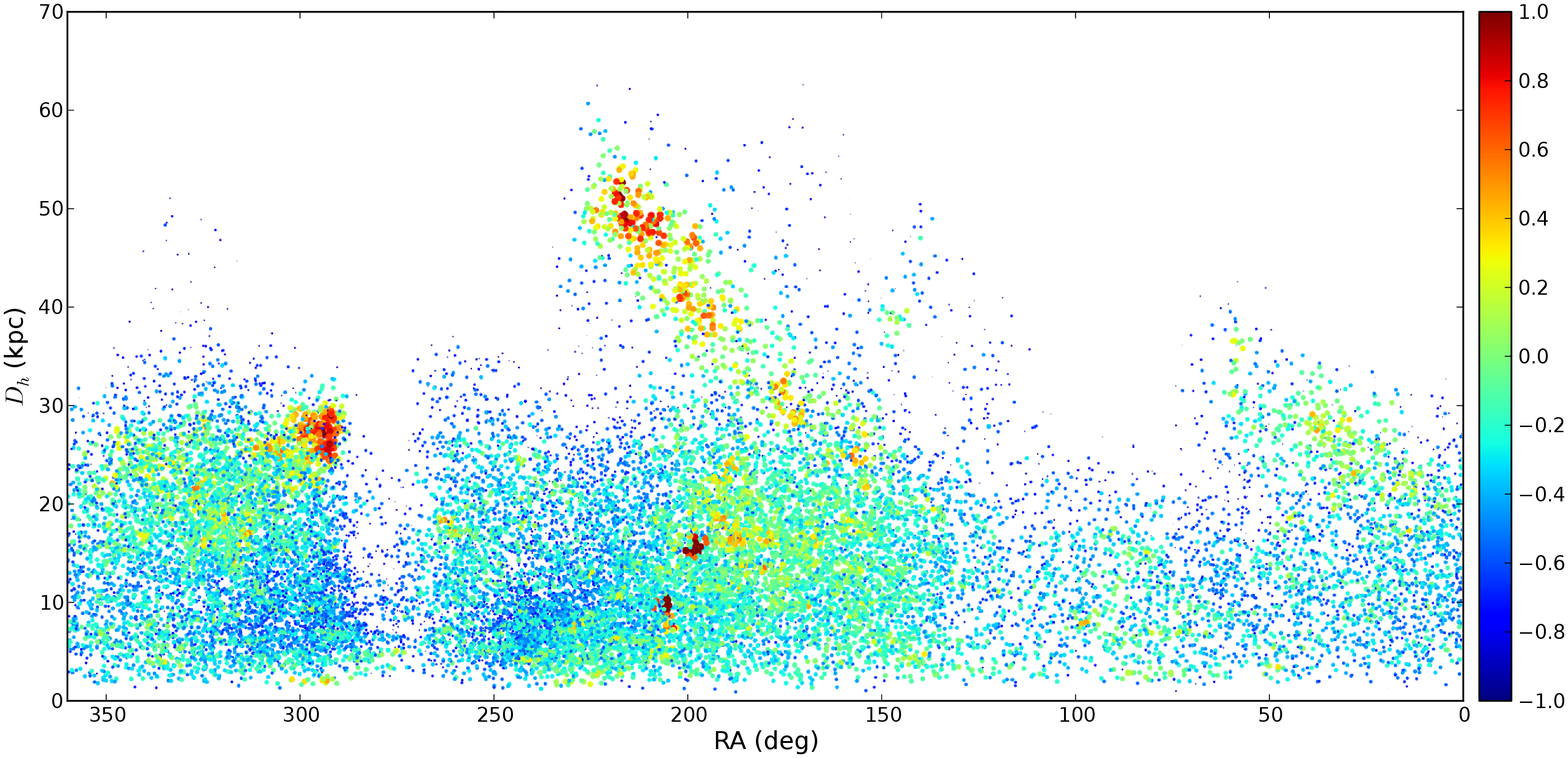}
    \caption{Map of RRab overdensity along the line-of-sight. This
      shows the density distribution of all Catalina RRab stars in the
      space spanned by RA and heliocentric distance, $D_{h}$, after the
      subtraction of a stellar halo model. The most prominent feature
      in the Figure is the Sagittarius stream. The color coding (see scale on the right) gives
      the logarithm of the ratio of the observed to the expected
      stellar density. To improve visualization, we saturate the color
      at $\pm 1$ and scale the size of the points according to the
      color. The position of the Sgr\,1 candidate substructure is at
      $\alpha \sim 300^{\circ}$, $D_{h} \sim 30$. It extends to
      $\alpha \sim 360^{\circ}$, where the distance gradient changes
      sign as the trailing tail starts to move away from the Sun until
      it disappears from the view at $\alpha\sim60^{\circ}$. In the
      north, the range of $220^{\circ} < \alpha < 150^{\circ}$ is
      dominated by the Sagittarius leading arm debris coming down from
      its apo-centre at around 55 kpc near $\alpha \sim 220^{\circ}$
      to $\sim$20 kpc at $\alpha\sim 150^{\circ}$. At lower distances,
      a more diffuse, cloud-like feature is the VOD. The
      two dense regions at $\alpha\sim 200^{\circ}$ correspond to M3 (NGC\,5272),
      at $\sim 10$ kpc, and M53 (NGC\,5024), at $\sim 15.5$ kpc.}
    \label{fig:fig14}
\end{figure*}

To visualize the overdensity and to highlight its connection to the
Sagittarius dwarf, Figure~\ref{fig:fig14} presents the density
distribution of all RRab stars detected by CRTS in the plane spanned
by right ascension and heliocentric distance, after subtracting a
model stellar halo. The RRab overdensities are colored according to
the logarithm of the ratio of the observed and model density,
$\log{(\rho_{\rm obs}/\rho_{\rm mod})}$. Here $\rho_{\rm obs}$ is the
observed local density defined by equation \ref{eq:SSTRUCT1}, and
$\rho_{\rm mod}$ is the model density defined by equation
\ref{eq:SSTRUCT2}. In this plotting scheme, the color is saturated at
1 and -1, and the size of the points is scaled to represent the
logarithm of the ratio of the densities in order to improve the
visualization.

According to previous studies \citep[e.g.,][]{smea03,sgrprec}, Sgr\,1,
the overdensity picked up in the South in the SSS dataset, is a
portion of the Sgr trailing stream. Correspondingly, the structure in
the north, visible in the CSS data, is the leading tidal tail. The
portion of the trailing stream found using the SSS data is located at
approximately $\alpha \sim 300^{\circ}$, $D_{h} \sim 30$ kpc. It is
curious to see that along the line of sight, the debris seem to spray
over a large range of distances. At the moment, it is difficult to
conclude with certainty whether this apparent clutter is a genuine
sub-structure in the Sgr trailing tail, or actually bits of other,
overlapping tidal streams like, for example, Cetus
\citep[see][]{Koposov2012}. Further away from the progenitor, there is
an obvious extension of the trailing debris to larger distances at
$0^{\circ} < \alpha < 70^{\circ}$. The leading tail runs from
$(\alpha,D_{h})=(230^{\circ},50~\rm{kpc})$ to
$(\alpha,D_{h})=(150^{\circ},20~\rm{kpc})$. Additionally, the Figure
shows a clump of stars at $\alpha \sim 150$ with distances between 30
and 50 kpc. This has already been reported as candidate D in
\citet{adea13b}, and, given its position and distance, is likely
associated with the Orphan stream \citep[see
  e.g.][]{orphan2007,SesarOrphan}.

\begin{figure*}
    \includegraphics[width=168mm]{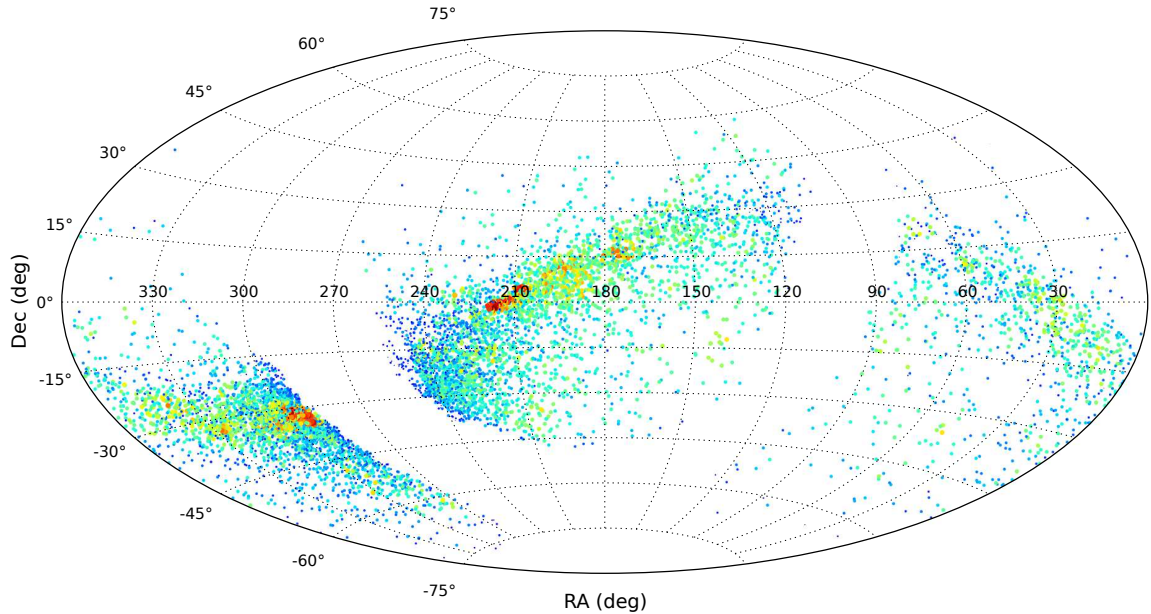}
    \caption{Distribution of RRab stars near the Sgr orbital plane in
      equatorial coordinates. This shows the Aitoff projection of the
      distribution of all stars with $\left|Z_{sgr}\right|<2.5$
      kpc. The color gives, according to the same scale used in Figure \ref{fig:fig14}, the ratio of the observed local density and
      the expected local density according to Equation
      \ref{eq:SSTRUCT2}. Note that the size of the points also scales
      with the density ratio. The scale and the continuity of the Sgr
      stream system is obvious, with the leading debris dominating the
      picture above the celestial equator and the trailing tail
      visible mostly in the south.}
    \label{fig:fig15}
\end{figure*}

Finally, Figure \ref{fig:fig15} displays the distribution of the RRab
stars around the Sagittarius plane in equatorial coordinates. Adopting
the \citet{smea03} Cartesian coordinate system, stars that are more
than 2.5 kpc away from the Sagittarius plane are removed. This
highlights the shape of the two main arms of the Sagittarius stream,
leading in the north, and trailing in the South. Sgr\,1, the
overdensity detected in this analysis (see Table~ \ref{tab:ovdc}), can
be seen at Dec $\sim -30^{\circ}$; note how this feature extends to
higher RA, from $\sim 300^{\circ}$ to $\sim 360^{\circ}$, as also seen
in Figure~\ref{fig:fig13a}.

\section{Conclusions}\label{sec:CONCL}

With the aim to find and classify RR Lyrae, we have carried out a
systematic study of a large variable star database covering most of
the Southern celestial hemisphere ($-75^{\circ} \leq \delta \leq
-22^{\circ}$). Using the data from the extension of the Catalina Sky
Surveys, the SSS, we have discovered more than $10,\!500$ ab-type RR~Lyrae, 
$\sim$90\% of which are new. Our sample covers a range in
apparent brightness from $10.5\leq V\leq 19.3$, corresponding to
heliocentric distances from $\sim 1$ to $\sim 50$ kpc.  The RRab stars
have been found using fully automated procedures that could be applied
to any other variability survey. These algorithms boast, depending on
the quality of the photometry, a completeness of RRab classification
greater than $60\%$ at $V=18$, and an efficiency of the period
estimation $>90\%$. The software leaves the possibility of manual
classification of the faintest stars, including other types of
periodic variables, using the lightcurve shape and period information
recorded.

We take advantage of the metallicity information available as a result
of the RRab lightcurve shape analysis to study the abundance patterns
in the Galactic stellar halo. The ${\rm [Fe/H]}$ distribution appears
roughly Gaussian with a mean metallicity of ${\rm [Fe/H]}=-1.4$~dex
and a dispersion of $0.3$~dex, which is consistent with previous
work. A comparison between the \citet{jjgk96} and the \citet{jnea13}
period-$\phi_{31}$-metallicity relation reveals clear differences at
both the metal-poor end, where the presence of a significant tail is
highlighted, as well as in the metal-rich range, where there seems to
be an excess of stars with metallicities between -1.0 and -0.5 dex,
according to estimates based on the \citet{jnea13} relation.

A spatial study of the metallicity gradients reveals a modest but
statistically significant ${\rm [Fe/H]}$ trend with both the
Galactocentric distance and the height above the disc plane. We
confirm that the RRab stars with lower metallicities are more common
further away from the Galactic center and the Galactic plane in
general. The radial gradient appears approximately flat in the first
$\sim7$~kpc but starts to steepen at larger distances. When compared
to the ${\rm [Fe/H]}$ behaviour with $z$, this flattening is not
present in the first few kpc. On the contrary, an accumulation of
high-metallicity RRab stars in the first $\sim2$~kpc is apparent,
which is consistent with the hypothesis that high-metallicity RRab
stars belong primarily to the Galactic disc population. This low-$z$,
high-metallicity concentration is even more pronounced when the
\citet{jnea13} ${\rm [Fe/H]}$ estimates are considered.

A preliminary analysis of possible halo sub-structure using the RRab
detected in the Southern hemisphere shows the presence of several
interesting overdensities. The locations and the structural properties
of the most significant twenty-seven of these are given in tabular
format. We have presented the properties of the most obvious
of the overdensity candidates discovered in our sample, the one with
the significance $>15\sigma$ and some $\sim$330 RRab associated. Given
its shape, spatial orientation, and position, this candidate, which we call Sgr~1,
is almost certainly associated with the Sagittarius stream system, in
particular, with the trailing arm. To fully confirm this, and to
compare our detection with the stream models available, it would be
beneficial to measure line-of-sight velocities of the RRab stars
within the bounds of the sub-structure. The second most significant
overdensity candidate, with a significance of $\sim 9~\sigma$, is
located near the Galactic center. It is stretched over many tens of
degrees in right ascension, which makes it difficult to pinpoint its
origin. Finally, we present all the overdensity candidates found in
our procedure with significances in excess of 2 $\sigma$. In order to
confirm -- or discard -- these candidates, a wide-area imaging and
spectroscopic survey is necessary. Our highly complete catalog of the
southern RR Lyrae fills in a large gap which has been so far
precluding a detailed and unbiased analysis of the Galactic stellar
halo, both in terms of its constituent sub-structure as well as its
shape and extent.


\section*{Acknowledgments} 

We thank A. K. Vivas, S. Duffau, and the anonymous referee, for useful comments
and discussions. Support for M.C. and G.T. is provided by the Ministry for the
Economy, Development, and Tourism's Programa Inicativa Cient\'ifica Milenio
through grant IC120009, awarded to the Millennium Institute of Astrophysics
(MAS); by Proyecto Basal PFB-06/2007; and by Proyecto FONDECYT Regular
\#1141141. Additional support for G.T. is provided by CONICYT Chile. CRTS and
CSDR1 are supported by the U.S.~National Science Foundation under grants
AST-0909182 and CNS-0540369. The SSS survey is funded by the National
Aeronautics and Space Administration under Grant No. NNG05GF22G issued through
the Science Mission Directorate Near-Earth Objects Observations Program.


\label{lastpage}

\end{document}